\documentclass[12pt]{article}

\usepackage{epsf,epsfig,epstopdf}
\usepackage{graphics}
\usepackage{amsmath,amssymb}
\usepackage{url}
\usepackage{slashed}
\usepackage{a4wide}

\usepackage[colorlinks=true, citecolor=blue, linkcolor=black, urlcolor=blue ]{hyperref}
\usepackage{cite}

\def\slash#1{#1 \hskip-0.45em /}

\newcommand{\dirac}[1]{\displaystyle{\not} #1}

\newcommand{\Eqn}[1]{Eq.~(\ref{#1})}

\begin{document}

\thispagestyle{empty}

\begin{flushright}
{\small
ITP-UU-13/04,
SPIN-13/02 \\
DESY 13-049 \\
PSI-PR-13-05, ZU-TH 06/13 \\
[0.2cm]
\today}
\end{flushright}

\vspace{\baselineskip}

\begin{center}

\vspace{0.5\baselineskip} \textbf{\Large\boldmath
Finite-width effects in unstable-particle \\[8pt]
production at hadron colliders
}
\\
\vspace{2\baselineskip}
{\sc P.~Falgari$^a$, A.S.~Papanastasiou$^b$, A.~Signer$^{c,d}$}\\
\vspace{0.5cm}
{\sl ${}^a$Institute for Theoretical Physics and Spinoza Institute,\\
Utrecht University, 3508 TD Utrecht, The Netherlands\\
\vspace{0.2cm}
${}^b$DESY, Deutsches Elektronen-Synchrotron,\\ Notkestra{\ss}e 85,
D-22607 Hamburg, Germany\\ 
\vspace{0.2cm}
${}^c$Paul Scherrer Institut,\\
CH-5232 Villigen PSI, Switzerland\\
\vspace{0.2cm}
${}^d$Institute for Theoretical Physics,\\
University of Zurich, CH-8057 Zurich, Switzerland\\
}

\vspace*{1.0cm}

\textbf{Abstract}\\

\vspace{1\baselineskip}

\parbox{0.9\textwidth}{We present a general formalism for the
  calculation of finite-width contributions to the differential
  production cross sections of unstable particles at hadron
  colliders. In this formalism, which employs an effective-theory
  description of unstable-particle production and decay, the matrix
  element computation is organized as a gauge-invariant expansion in
  powers of $\Gamma_X/m_X$, with $\Gamma_X$ and $m_X$ the width and
  mass of the unstable particle.  This framework allows
  for a systematic inclusion of off-shell and non-factorizable
  effects whilst at the same time keeping the computational effort minimal
  compared to a full calculation in the complex-mass scheme.  As a
  proof-of-concept example, we give results for an NLO calculation of
  top-antitop production in the $q \bar{q}$ partonic channel. As
  already found in a similar calculation of single-top production, the
  finite-width effects are small for the total cross section, as
  expected from the na\" ive counting $\sim \Gamma_t/m_t \sim 1\%$. 
  However, they can be sizeable, in excess of $10\%$, close to edges of
  certain kinematical distributions. The dependence of the results on the mass
  renormalization scheme, and its implication for a precise extraction
  of the top-quark mass, is also discussed.}

\end{center}

\newpage

\section{Introduction}

The performance of the Large Hadron Collider (LHC) so far has been
extremely successful, with about $5 \text{fb}^{-1}$ of integrated
luminosity collected in the 2011 run and more than $20\,
\text{fb}^{-1}$ in 2012.  This has led to an unprecedented accuracy in
the measurements of many Standard Model (SM) cross sections and
distributions, both in the electroweak and strong sectors, and also
to the discovery of a new particle consistent with a SM Higgs
boson. With the high-energy run which will follow the 2013
shutdown, experimental errors are bound to reduce even further,
leading to higher-precision measurements and to the possible
observation of new-physics effects beyond the SM. This high 
experimental precision clearly motivates similarly
accurate theoretical predictions for cross sections and kinematical
distributions, so that on the one hand the clean extraction of signals 
from the data is possible and, on the other hand, contributions 
to the backgrounds to processes of interest can be accurately constrained.
           
Most of the phenomenologically interesting processes at the LHC, such
as $W$ and $Z$ boson production, top-quark production and Higgs
production, not to mention beyond-the-Standard-Model (BSM) processes,
like supersymmetric (SUSY) particle production, involve massive
unstable particles. These particles are not asymptotic states and
show up in detectors as energetic jets and leptons, often being 
accompanied by large transverse missing energy $\dirac{E}_T$.
The necessity of precise theoretical predictions therefore raises the
question of how to correctly treat the decay of the intermediate
unstable particle to the physical final states. For observables which
are inclusive in the final states originating from the
unstable-particle decay, it is often sufficient to treat the massive
particles as stable, ignoring their decay. The error associated with
this approximation is formally of order $\Gamma_X/m_X$, where 
$\Gamma_X$ and $m_X$ are the width and mass of the unstable particle. 
For the aforementioned processes this corresponds to less than a few
percent, i.e. typically smaller than the experimental errors. 
Clearly, while the stable approximation is appropriate for the
total cross section, it cannot be used to predict arbitrary
kinematical observables. Moreover, the error associated with this
approximation could actually be numerically sizeable for
new, yet undiscovered wide resonances, for example strongly-decaying 
SUSY particles.

A step forward towards a realistic description of production and decay
of an unstable particle $X$ is the narrow-width approximation (NWA)
which is a framework commonly used in the context of high-energy
calculations for hadron colliders. In the NWA the particle is produced
and allowed to decay to the physical final states while remaining on
shell.  At next-to-leading order (NLO), radiative corrections are
given by factorizable virtual and real contributions to the on-shell
production and decay subprocesses.  While technically only slightly
more involved than the stable-top approximation, the NWA preserves
spin correlations between the production and decay subprocesses, and
allows for realistic kinematical cuts on the momenta of the physical
final states (i.e. leptons and jets).

The NWA includes neither off-shell effects related to
the virtuality of the intermediate unstable-particle propagator, nor
non-factorizable corrections linking the production and decay subprocesses.  
Sub-resonant or non-resonant contributions, which correspond to diagrams 
with the correct physical final state but which involve fewer or no 
intermediate unstable-particle propagators, are also neglected. 
As in the stable approximation, these finite-width effects
are expected to be small, of order $\Gamma_X/m_X$, for
inclusive-enough observables. This is due to large cancellations between
virtual and real non-factorizable corrections and also because of the
suppression of non-resonant contributions.  However, for arbitrary kinematical 
distributions and in particular, close to certain kinematical thresholds where the 
cancellations mentioned above are less effective, finite-width effects can be large.  
Strikingly, in Refs.~\cite{Berdine:2007uv,Kauer:2007zc,Uhlemann:2008pm} it was pointed out
that the na\"ive expectation of the error associated with the NWA can be underestimated  
by an order of magnitude for BSM processes where the mass of daughter
particles approaches the mass of the parent particle $X$. This is
relevant for searches of SUSY in decay cascades, where one often
observes some degree of mass degeneracy between particles in different
steps of the cascade. Recently, non-negligible off-shell effects
were observed even in Higgs production and decay to massive vector
bosons \cite{Kauer:2012hd}, due to interferences between resonant and
non-resonant contributions. Thus, it is clear that an approach that goes beyond 
the NWA and which includes at least the dominant finite-width effects is desirable.

A possible solution to the issue of finite-width effects is clearly
the calculation of the full, gauge-invariant set of diagrams
corresponding to a given physical final state. This approach includes the
coherent sum of resonant and non-resonant contributions, treats the
intermediate resonant particles as fully off-shell and contains both
factorizable and non-factorizable corrections at NLO.  Self-energy
contributions can be resummed in the unstable-particle propagator in a
consistent gauge-invariant way using, for example, the complex-mass
scheme \cite{Denner:2005fg,Denner:2006ic}. Examples applying 
the complex-mass scheme to production of unstable particles at NLO include
the calculations of four-fermion production at an $e^+ e^-$ collider
\cite{Denner:2005fg}, Higgs decay to vector-boson pairs
\cite{Bredenstein:2006ha} and two recent independent calculations of
off-shell effects in $t \bar{t}$ production \cite{Denner:2010jp,
  Bevilacqua:2010qb,Denner:2012yc}.  While the complex-mass scheme approach is
completely general and very flexible, allowing the calculation of
arbitrary kinematical distributions, the full NLO computation is
technically challenging, requiring both the calculation of a much larger
set of diagrams than for the corresponding on-shell process and
special techniques to handle 5- or 6-point functions with complex
masses.

An alternative approach to the full NLO calculation was presented in
Ref.~\cite{Falgari:2010sf} and applied to processes of $t$-channel and
$s$-channel single-top production \cite{Falgari:2010sf,Falgari:2011qa}.  
The approach of Ref.~\cite{Falgari:2010sf} is the generalization of the effective
field theory (EFT) description of \emph{resonant}-particle production
of Ref.~\cite{Beneke:2004km}, which was employed in the calculation of
inclusive $W$-pair production at an $e^+ e^-$ collider
\cite{Beneke:2007zg,Actis:2008rb}.  The EFT calculation results in a
systematic, gauge-invariant expansion of the matrix elements in powers
of $\Gamma_X/m_X$, in a way which can be considered a generalization
of the pole approximation \cite{Stuart:1991xk,Aeppli:1993rs}. Compared
to the full NLO calculation in the complex-mass scheme, the
effective-theory approach has the advantage of identifying the terms
that are relevant to achieving a given target accuracy prior to the actual
calculation. This greatly reduces the complexity of the computation
while at the same time allows for the inclusion of the leading off-shell
and non-factorizable effects in a completely differential manner. 
For single-top production finite-width effects were found
to be small for inclusive-enough observables, although they can be large
close to the kinematical edges of some distributions. This general
picture is consistent with the results found by the full NLO
calculations of top-pair production \cite{AlcarazMaestre:2012vp}.

In this paper we give a second example of the application of the EFT
formalism of Ref.~\cite{Falgari:2010sf} and calculate the cross
section for the pair-production process $q \bar{q} \rightarrow t
\bar{t} \rightarrow W^+ W^- b \bar{b}$. The top-pair production
process has been studied extensively over the years in the stable-top
and narrow-width approximations and, more recently, using the
complex-mass scheme (a detailed list of references is given in Section
\ref{sec:amplitudes}). It thus represents a perfect proof-of-concept
calculation by which to test the validity of the EFT formalism, extend
it to more than one unstable particle and to compare it to different
available approaches.  The paper is organized as follows: in Section
\ref{sec:EFT} we review the effective-theory formalism and introduce a
treatment of real corrections which differs slightly from the one used
in Ref.~\cite{Falgari:2010sf}.  The calculation of the LO and NLO
relevant amplitudes for the specific example of $t \bar{t}$ production
is described in Section \ref{sec:amplitudes}.  In Section
\ref{sec:res} we present results for several distributions and assess
the effect of finite-width contributions by comparing the
effective-theory predictions with results obtained in the NWA. In that
same section we also discuss the effects of using different
mass-renormalization schemes (more precisely the pole and PS schemes)
on the cross section and distributions, and their possible
implications for a precise extraction of the top-quark mass from
data. Finally, our conclusions are given in Section
\ref{sec:conclusions}.

\section{Effective-theory description of unstable-particle production}
\label{sec:EFT}

The effective-theory framework for the description of
unstable-particle production used in this work was first formulated
for the total cross section in Ref.~\cite{Beneke:2004km} and applied
to the case of inclusive $W^+ W^-$ production at an $e^+ e^-$ collider
in Refs.~\cite{Beneke:2007zg,Actis:2008rb}. The formalism was later
extended to the more general case of differential cross sections and
applied to single-top production at hadron
colliders~\cite{Falgari:2010sf,Falgari:2011qa}. In this section we
review the main features of this approach, referring the reader to the
aforementioned references for further details.

Unstable-particle effective theory is built upon the hierarchy of two
scales, namely, the typical virtuality of the resonant unstable
particle $X$, which is set by its decay width, $p_X^2-m_X^2 \sim m_X
\Gamma_X$, and the particle mass $m_X$.  This hierarchy is encoded in
the ratio $\Gamma_X/m_X \ll 1$. The latter is treated as a small
parameter, $\delta$, in line with the strong and electroweak coupling
constants $\alpha_s$, and $\alpha_{ew}$ and allows for a systematic
expansion of the full matrix elements.  These different expansion
parameters are related by the counting scheme
\footnote{Note that in (\ref{eq:counting}) we assume
  that the unstable-particle decay proceeds via electroweak decay
  channels, $\Gamma_X \propto \alpha_{ew}$. For resonances decaying
  via strong interactions the counting scheme is $\Gamma_X \sim
  \alpha_s \sim \sqrt{\alpha_{ew}}$.}
\begin{equation}\label{eq:counting}
\frac{\Gamma_X}{m_X} \sim \alpha_s^2 \sim \alpha_{ew} \, .
\end{equation} 
In the following we will generically refer to any of the above
parameters as $\delta$.  The expansion in
$\delta$ is implemented at the Lagrangian level, replacing the (B)SM
fields and interactions with effective fields and vertices. The
effective fields are associated with different momentum
regions defined according to the scaling of their momenta
with respect to the parameter $\delta$ and encode the physics at the
two very different scales that characterize the production and decay
process\footnote{Here and in the following we assume that the
  invariants $s_{i j}=2 p_i \cdot p_j$ constructed from the external
  momenta are of the same order of $m_X^2$, and we treat them as a
  single scale.}. For the problem at hand these momentum regions are a
\emph{hard} region ($q_0 \sim |\vec{q}| \sim m_X$), a \emph{soft}
region ($q_0 \sim |\vec{q}| \sim m_X \delta$) and \emph{collinear}
regions ($n_i \cdot q \sim m_X \delta, \, \bar{n}_i \cdot q \sim m_X,
\, q_\perp \sim m_X \sqrt{\delta}$). Here $n_i, \,\bar{n}_i$ are
light-like vectors associated with the momenta of the external
massless particles, $n_i=(1,\vec{p}_i/|\vec{p}_i|)$,
$\bar{n}_i=(1,-\vec{p}_i/|\vec{p}_i|)$, and $q_\perp$ is the
remaining, perpendicular component of the momentum $q$.
 
In the effective theory only low-virtuality modes with $p^2 \lesssim
m_X^2 \delta$ are kept as dynamical degrees of freedom, and are
described by effective fields in the Lagrangian. These include, in
particular, a field $\Phi_X$ to describe the resonant unstable
particle $X$. The finite-width of the particle is resummed into the
leading EFT kinetic term in a generalisation of the heavy-quark
effective theory (HQET) Lagrangian in the case of a non-vanishing
width \cite{Beneke:2004km},
\begin{equation} \label{eq:kinET}
{\cal L}^{(0)}_{\text{EFT}, kin}=2 \hat{m}_X \Phi^\dagger_x \left(i v
\cdot \partial-\frac{\Omega_X}{2} \right) \Phi_X \, , 
\end{equation}   
where $\hat{m}_X v$, with $v^2= 1$, represents the large, on-shell
component of the resonant-particle momentum, and $\hat{m}_X$ is the
renormalized mass in a generic renormalization scheme.  The
coefficient $\Omega_X$ is related to the complex pole $\mu_X^2 \equiv
m_X^2-i m_X \Gamma_X$ of the full unstable-particle propagator,
\begin{equation} \label{eq:omega}
\Omega_X=\frac{\mu_X^2-\hat{m}_X^2}{\hat{m}_X} \, .
\end{equation}
In the pole scheme, $\hat{m}_X=m_X$, $\Omega_X$ has the simple form
$\Omega_X=-i \Gamma_X$. Additional terms in the EFT Lagrangian are
given by bilinear terms for soft and collinear fields,
power-suppressed corrections to \eqref{eq:kinET} and terms describing the interaction of
$\Phi_X$ and collinear fields with soft fields. Hard modes are not explicitly
part of the effective Lagrangian and their contribution is encoded in
\emph{matching coefficients} multiplying effective interaction
vertices. These can be schematically parameterized as
\begin{equation} \label{eq:EFT_interactions}
{\cal C}_{i,P}(\mu_X) {\cal F}_P^i(\Phi^\dagger_X,\phi_c,\phi_s,\partial_\mu), \hspace{5mm}  
{\cal C}_{j,D}(\mu_X) {\cal
  F}_D^j(\Phi_X,\phi_c,\phi_s,\partial_\mu), \hspace{5mm} 
 {\cal C}_{k,NR}(\mu_X) {\cal
  F}_{NR}^k(\phi_c,\phi_s,\partial_\mu)
\, , 
\end{equation}
where ${\cal F}_P^i$, ${\cal F}_D^j$ and ${\cal F}_{NR}^k$ denote
functions of fields and derivatives and the indices $i, \, j, \, k$
label different Lorentz structures. $\phi_{c,s}$ generically represent collinear
and soft fields and ${\cal C}_{i,P}$ and ${\cal C}_{j,D}$ are the hard
matching coefficients of the production and decay effective vertices,
which are computed from on-shell SM amplitudes. In this context
``on-shell" has to be understood as $p_X^2=\hat{m}_X^2+\hat{m_X}
\Omega_X=\mu_X^2$, meaning that the effective couplings in the
Lagrangian are in general complex. This is a feature that the EFT
framework shares with the complex-mass scheme.  The interaction terms ${\cal C}_{k,NR}
{\cal F}_{NR}^k$ encode the contribution of non-resonant
configurations which also contribute to the cross section starting from a
certain order in $\delta$.  Note that in order to describe
pair-production of unstable particles, e.g. $t \bar{t}$ production,
two distinct resonant fields, $\Phi_t$ and $\Phi_{\bar{t}}$, have to be 
introduced which annihilate a top and an antitop state respectively. 
Furthermore, the kinetic terms will contain two velocities $v$
and $\bar{v}$ which are generally different.

\subsection{Born amplitudes}

\begin{figure}
\begin{center}
\includegraphics[width=0.25 \linewidth,angle=-90]{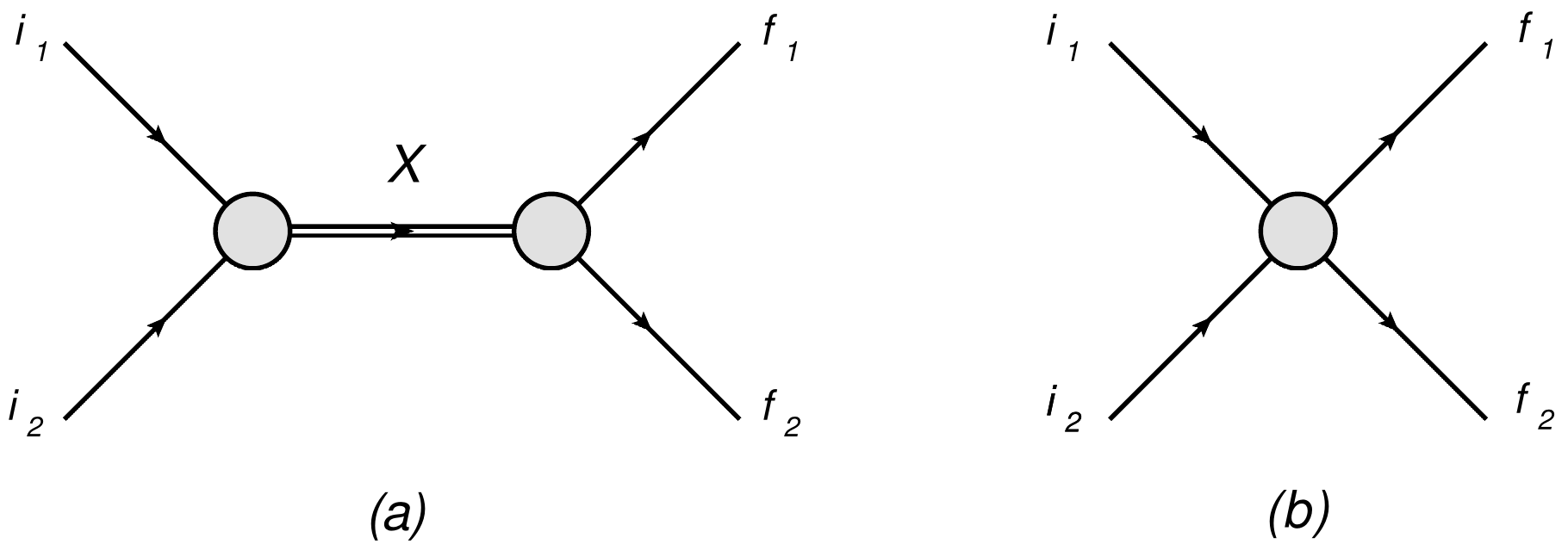}
\end{center}
\caption{Feynman-diagram topologies contributing to the process $i_1
  i_2 \rightarrow f_1 f_2$. (a) resonant production through an
  intermediate unstable $X$; (b) non-resonant production.}
\label{fig:proc_tree}
\end{figure}
From a practical point of view, at tree-level the EFT result
coincides with an expansion of the matrix element around the complex
pole of the propagator. In fact, the EFT approach can be viewed as a
generalization, to arbitrary order in $\delta$, of the pole
approximation \cite{Stuart:1991xk,Aeppli:1993rs}. To be specific, let
us consider the process
\begin{equation} \label{eq:process}
i_1(p_1) i_2(p_2) \rightarrow X(p_X),\,...  
\rightarrow f_1(k_1) f_2(k_2) \, ,
\end{equation} 
where the final-state particles $f_1$ and $f_2$ can originate from the
decay of an intermediate unstable scalar particle $X$ (the
generalization to higher spins is trivial), or from other production
mechanisms. The Feynman-diagram topologies contributing to this
process at the tree-level are given by the resonant and non-resonant
contributions drawn in Figure~\ref{fig:proc_tree}, where the grey
blobs denote the model-dependent production and decay vertices. The
tree-level amplitude can be written as
\begin{equation} \label{eq:ampl_tree}
{\cal A}_{\text{tree}} = \frac{{\cal V}_P(\{p_i\},p_X) 
{\cal V}_D(p_X,\{k_i\})}{p_X^2-m_{0,X}^2} 
+ {\cal N}(\{p_i\},\{k_i\}) \, , 
\end{equation}
where ${\cal V}_P$ and ${\cal V}_D$ represent the vertices for the
production and decay of the (off-shell) particle $X$ respectively,
while ${\cal N}$ contains the non-resonant contributions. $m_{0,X}$
denotes the bare mass of the particle $X$. It is immediately clear
from \Eqn{eq:ampl_tree} that due to the intermediate resonant
propagator, $p_X^2-m_{0,X}^2 \sim m_X^2 \delta$, only resonant
diagrams will contribute to the amplitude at leading order in
$\delta$, while ${\cal N}$ will generally be suppressed by extra
powers of $\delta$.  The actual suppression of the non-resonant
contributions is determined from the interplay between the suppression 
from the missing resonant propagators and the scaling of the couplings appearing
in ${\cal V}_{P}$, ${\cal V}_{D}$ and ${\cal N}$, see \Eqn{eq:counting}.

The gauge-invariant expansion in $\delta$ of the amplitude,
\Eqn{eq:ampl_tree}, is obtained by expanding the matrix elements for
production and decay around $p_X^2=\mu_X^2$. This requires a
projection of the external momenta $\{p_i\}$, $\{k_i\}$ onto on-shell
configurations $\{\bar{p}_i\}$, $\{\bar{k}_i\}$, with
\begin{equation}\label{eq:proj}
\bar{p}_i =\bar{p}_i(p_i,p_X) \hspace{2 cm} \bar{k}_i
=\bar{k}_i(k_i,p_X) \, .
\end{equation} 
The on-shell projection is chosen such that momentum is exactly
conserved at each vertex and $\bar{p}_X^2=\mu_X^2$.  Note that the
explicit form of the projection in \Eqn{eq:proj} is not unique, being
defined up to terms of order $\delta$. However, calculations at order
$\delta^n$ obtained with different projections, deviate from each other
by sub-leading corrections of order $\delta^{n+1}$, i.e. always an
order higher than the target accuracy of the calculation.

Adopting \Eqn{eq:proj}, the expansion in $\delta$ of the tree-level
matrix element reads
\begin{eqnarray} \label{eq:ampl_tree_exp}
{\cal A}_{\text{tree}} &=& \frac{{\cal A}_P {\cal
    A}_D}{\Delta_X}\nonumber\\ &&+\frac{1}{\Delta_X} \left((p_i-\bar{p}_i)
\frac{\partial {\cal V}_P}{\partial p_i} {\cal A}_D+{\cal A}_P
\,(k_i-\bar{k}_i) \frac{\partial {\cal V}_D}{\partial k_i} \right) +{\cal N}
+ \ldots
\end{eqnarray}
where ${\cal A}_{P} \equiv {\cal V}_{P}(\{\bar{p}_i\},\bar{p}_X)$,
${\cal A}_{D} \equiv {\cal V}_{D}(\bar{p}_X,\{\bar{k}_i\})$ are
evaluated with the projected momenta $\bar{p}_i,\,\bar{k}_i$, and the
ellipses represent higher-order terms in $\delta$. As expected, the
leading-order amplitude (first line in \Eqn{eq:ampl_tree_exp})
corresponds to the resonant diagram with the vertices for
production and decay of the unstable particle replaced by
(gauge-invariant) on-shell amplitudes. These amplitudes are in fact
directly related to the matching coefficients $C_{0,P}$, $C_{0,D}$ of
the (leading) effective vertices appearing in
\Eqn{eq:EFT_interactions}.  The modified propagator $\Delta_X \equiv
p_X^2-\mu_X^2=p_X^2-\hat{m}_X^2-\hat{m}_X\Omega_X$ Dyson-resums the
finite-width effects related to the self-energy $\Pi_X(p_X^2)$ of the
particle $X$:
\begin{equation} \label{eq:prop_soft}
\frac{1}{p_X^2-\hat{m}_X^2+\Pi_X(p_X^2)} = \frac{1}{\Delta_X}+\ldots
\end{equation} 
with higher-order terms in $\delta$ indicated by the ellipses.  Note
that it is not the full self-energy that is kept in the propagator. 
Only the gauge-invariant hard part of the self-energy
$\Omega_X$ contributes to the matching coefficient and is resummed in the
propagator. The sub-leading gauge-violating residual soft terms are
included perturbatively and are combined with other contributions to
form a separate gauge-invariant part of the one-loop amplitude.  The
next term in $\delta$ in the expansion of the amplitude (second line
in \Eqn{eq:ampl_tree_exp}) receives contributions from non-resonant 
diagrams, as expected, but also from resonant ones in which the
propagator $\Delta_X$ is cancelled by terms of the form $p_i-\bar{p}_i
\sim \delta$ and $k_i-\bar{k}_i \sim \delta$, which originate from the expansion
around on-shell configurations. 
In the EFT language
these contributions are described by effective four-particle
operators, i.e. terms of the form ${\cal C}_{k,NR} {\cal F}_{NR}^k$ in 
\Eqn{eq:EFT_interactions}.  We stress that while the
leading resonant term is gauge-invariant, at higher orders in $\delta$
only the sum of resonant and non-resonant contributions is on the whole 
gauge independent.

\subsection{Virtual corrections}
\label{sec:EFT_virtual}

\begin{figure}
\begin{center}
\includegraphics[width=0.75 \linewidth]{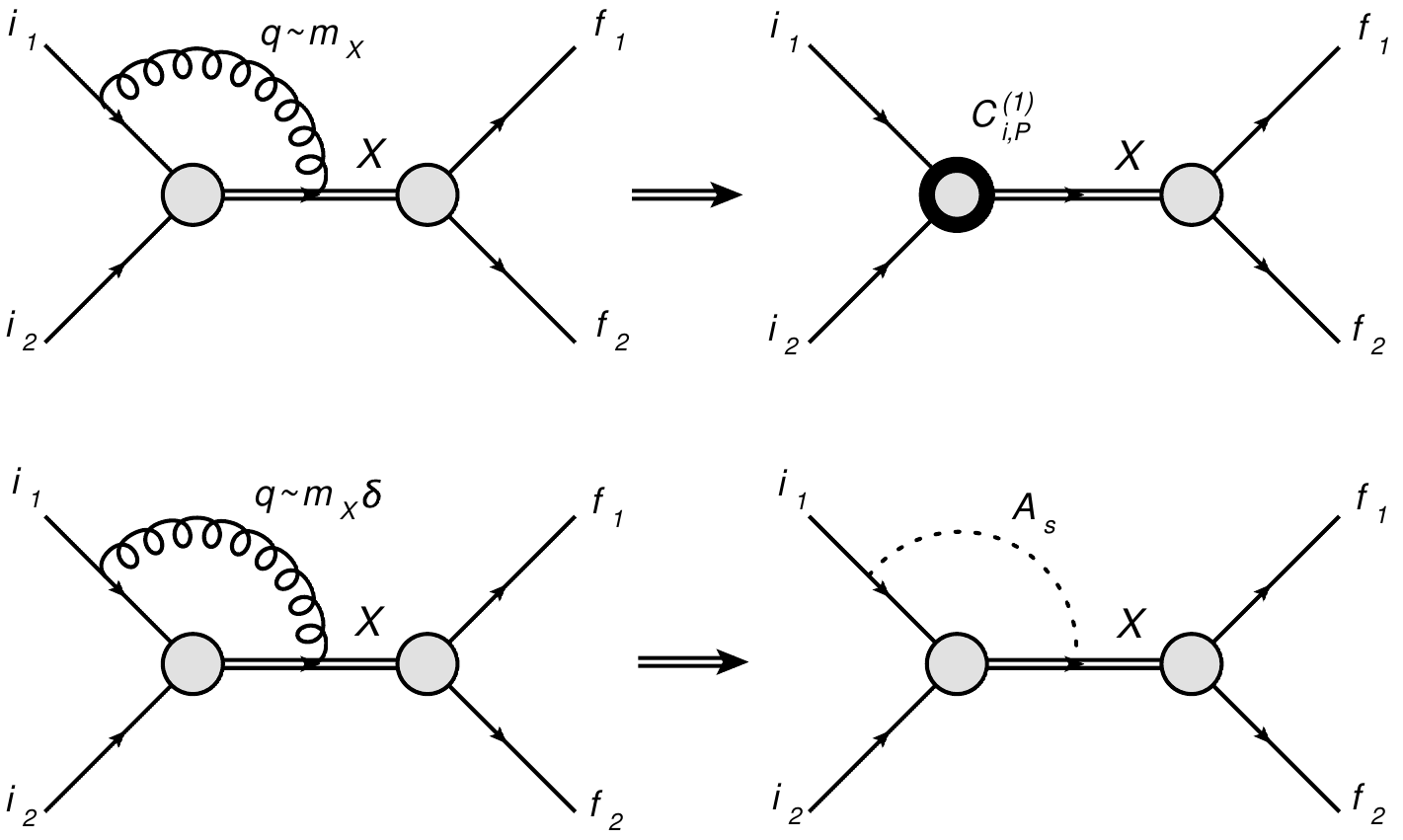}
\end{center}
\caption{Correspondence between the expansion by regions and the EFT
  calculation: the hard-region contribution (top left) corresponds to
  a ${\cal O}(\alpha_s)$ correction to the production matching
  coefficient $C_{i,P}$ (top right), while the soft-region
  contribution (bottom left) reproduces one-loop soft-gluon
  corrections in the EFT (bottom right).}
\label{fig:proc_regions}
\end{figure}
Virtual corrections in the EFT framework are divided into two
categories: loop corrections to the EFT matrix elements, where the
degrees of freedom flowing in the loops are given by the effective
fields $\Phi_X, \, \phi_c, \phi_s$, and corrections to the hard
matching coefficients $C_{i,P}, \, C_{j,D}, \, C_{k,NR}$. The latter
are computed by a ``matching procedure'' which arises by 
requiring that the full-theory matrix element and the EFT result
coincide order by order in $\delta$. They can be written as a power
series in the couplings (the QCD coupling, in this case),
\begin{equation}
C_{i,P} = C_{i,P}^{(0)}+\frac{\alpha_s}{2 \pi} C_{i,P}^{(1)}+...
\end{equation}
with similar expressions for $C_{j,D}$ and $C_{k,NR}$. In practice,
the EFT matrix elements and the matching coefficients can be obtained
from an expansion of SM matrix elements using the method of regions
\cite{Beneke:1997zp}, which we find to be more convenient here than an
explicit two-step calculation of matching coefficients and
effective-theory matrix elements.  This is the approach we adopt for
the calculation of the amplitudes for $t \bar{t}$ production in
Section~\ref{sec:amplitudes}, where details on how the expansion by
regions is implemented for the specific case of resonant top-pair
production are given.

As already outlined, the momentum regions relevant to the expansion are the 
hard, soft and collinear regions. The contributions of hard
momenta ($q \sim m_X$) correspond to the matching coefficients,
while those from an expansion in the soft region ($q \sim m_X \delta$) of 
the full SM integrals reproduces loop contributions in the effective theory.  
This is schematically depicted in Figure~\ref{fig:proc_regions} for the
case of the one-loop gluonic correction to the production
vertex.  Hard corrections correspond to factorizable contributions to
the production or decay subprocesses \cite{Chapovsky:2001zt}, and, at
leading order in $\delta$, coincide with the matrix element for
production or decay of the on-shell massive particle(s).  On the other
hand, soft corrections encode non-factorizable interferences between
the production and decay subprocesses as well as off-shell effects.  Note
that in general the contribution from collinear regions is needed to reproduce the full
SM matrix element. However, for the case considered, this contribution vanishes if loop 
integrals are regularized dimensionally and the external light-fermion 
masses are set to zero, as done in this work. It can thus be safely ignored in
the following discussion.  In the soft region, the unstable-particle propagator
in the loop, $(p_X-q)^2-m_X^2 \sim m_X^2 \delta$, is resonant and has
to be Dyson-resummed,
\begin{equation}
\frac{1}{(p_X-q)^2-\hat{m}_X^2+\Pi_X((p_X-q)^2)} 
= \frac{1}{\Delta_X-2 \bar{p}_X \cdot q}+... \, ,
\end{equation} 
with the ellipses denoting, as usual, higher-order terms in $\delta$.
As in the tree-level matrix element, only the on-shell gauge-invariant
part of the self energy is resummed in the resonant propagator. In the
hard region, where $(p_X-q)^2-\hat{m}_X^2 \sim \hat{m}_X^2 \gg m_X
\Gamma_X$, no self-energy resummation is necessary and
gauge-invariance is similarly preserved.
   
\begin{figure}
\begin{center}
\includegraphics[width=0.25 \linewidth,angle=-90]{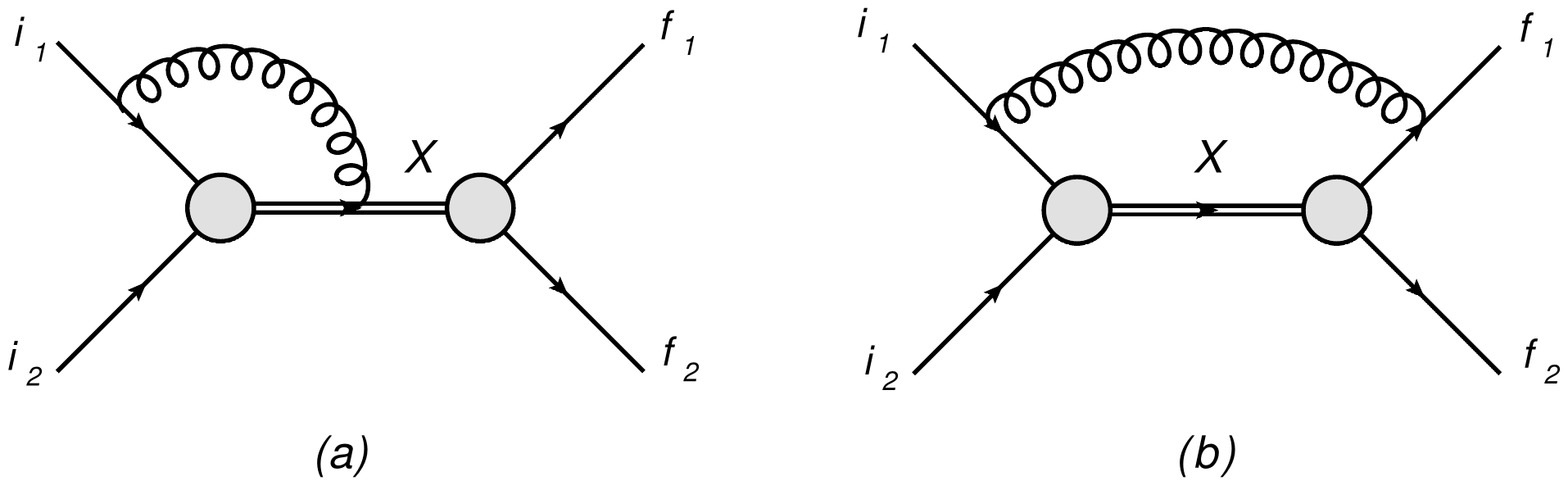}
\end{center} 
\caption{Examples of triangle and box diagrams contributing to the
  production and decay of the scalar $X$ at one loop.}
\label{fig:scaling_ex}
\end{figure} 
An important feature of the effective-theory approach is
that the EFT counting scheme, Eq.~(\ref{eq:counting}), enables one to
assign the correct parametric scaling to any particular contribution
of a Feynman diagram prior to its actual calculation. This determines
whether or not it has to be computed to obtain a given accuracy in
$\delta$. 

Suppose for example that one wants to compute all terms
which scale as $\alpha_s \sim \delta^{1/2}$ relative to the leading
tree-level contribution. From now on we will refer to such corrections as
NLO corrections. If we ignore the couplings appearing in the production and decay
vertices ${\cal V}_P$ and ${\cal V}_D$, the leading tree-level matrix
element (first term in \Eqn{eq:ampl_tree_exp}) scales as $1/\Delta_X
\sim \delta^{-1}$, and NLO is thus defined as the sum of all terms
scaling as $\delta^{-1} \delta^{1/2}=\delta^{-1/2}$. One can then
immediately see that the sub-leading tree-level terms in
\Eqn{eq:ampl_tree_exp}, which scale as $\delta^{-1} \delta=1$
(assuming that ${\cal V}_P {\cal V}_D \sim {\cal N}$), are not part of
the NLO matrix element, contributing to the amplitude only at
NNLO. Consider now the triangle and box integrals shown in
Figure~\ref{fig:scaling_ex}. In the soft region of the loop momentum
($q \sim m_X \delta$), the unstable-particle propagator inside the
loop scales as $\delta^{-1}$, while the gluon and fermion propagators
scale as $\delta^{-2}$ and $\delta^{-1}$ respectively. Taking into
account the scaling of the volume element $d^4 q \sim \delta^4$, one
finds the scaling $\delta^{-1} \alpha_s \delta^4 \delta^{-2}
\delta^{-1} \delta^{-1} \sim \delta^{-1/2}$ for the soft part of the
triangle, while the soft part of the box scales as $\alpha_s \delta^4
\delta^{-2} \delta^{-1} \delta^{-1} \delta^{-1} \sim \delta^{-1/2}$.
Therefore, the soft limit of both the triangle and box contribute at
NLO. In the hard region ($q \sim m_X$) all the propagators inside the
loop scale parametrically as $\sim 1$ (in units of the mass
$m_X$). Thus, the hard part of the triangle scales as $\delta^{-1}
\alpha_s 1^4/1^3 \sim \delta^{-1/2}$, while for the box one finds
$\alpha_s 1^4/1^4 \sim \delta^{1/2}$. In this case only the hard part
of the triangle counts as NLO, while the hard box integral is highly
suppressed, scaling as a N$^3$LO correction.  Given that for the
simple process (\ref{eq:process}), the hard box in
Figure~\ref{fig:scaling_ex}(b) is the most complicated integral at one
loop, discarding it simplifies the NLO calculation.  As we will see in
Section~\ref{sec:virtual}, for the phenomenologically relevant case of
top-pair production one can use the same scaling arguments to show
that all one-loop hard 5- and 6-point functions are parametrically
suppressed and need not be computed to achieve NLO accuracy in the
resonant region.

\subsection{Real corrections}
\label{sec:EFT_real}

Even though the expansion by regions is well understood for loop
diagrams, it is less clear how the expansion can in general be
implemented for real corrections, since in the presence of an extra
massless particle with momentum $q$, the expansion parameter can be
$p_X^2-\mu_X^2$, $(p_X+q)^2-\mu_X^2$ or both. While the split into
hard and soft contributions is possible for the total cross section,
by relating real corrections to cut one-loop diagrams via the optical
theorem, the formulation of the expansion in hard and soft
contributions is not straightforward for an arbitrary observable. In
Refs.~\cite{Falgari:2010sf,Falgari:2011qa} we proposed a way to
circumvent this problem by using the exact real-radiation matrix
element and expanding the integrated infrared and 
collinear subtraction terms in $\delta$ so as to properly match the 
singularity structure of the expanded virtual matrix element. 

While this ``ad-hoc" treatment of the real matrix element 
was sufficient to illustrate the basic features of the EFT
approach and to assess the contribution of non-factorizable
corrections in single-top
production, a theoretically more satisfactory treatment of real
corrections is certainly desirable. In particular, one would wish for a
framework in which a strict expansion in $\delta$ for both virtual and
real corrections is achieved and gauge-invariance is exact
order-by-order in $\delta$. Here we introduce such a framework which
will be used in Section \ref{sec:real} for the calculation of real
corrections to $t \bar{t}$ production.

\begin{figure}
\begin{center}
\includegraphics[width=0.95 \linewidth,angle=0]{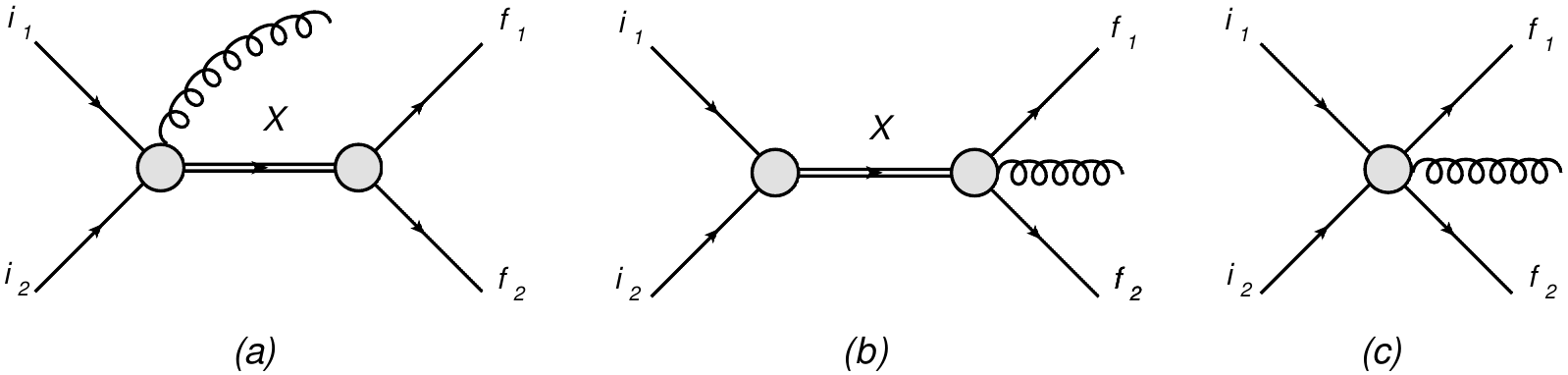}
\end{center} 
\caption{Real gluonic corrections to the process $i_1 i_2 \rightarrow f_1 f_2$.}
\label{fig:X_real}
\end{figure} 
Real gluon-emission contributions fall into three categories, as indicated
in Figure~\ref{fig:X_real}. Topology (a) and (b) represent emission
from the production and decay subprocess respectively, whereas topology
(c) is due to generic non-resonant real-emission. Analogously 
to \Eqn{eq:ampl_tree_exp}, the corresponding amplitude can be
written as
\begin{equation} \label{eq:A_real} 
{\cal A}_\text{real} = \frac{{\cal A}_{P+g} {\cal A}_D}{\Delta_X}
+\frac{{\cal A}_{P} {\cal A}_{D+g}}{\Delta_{Xg}} + {\cal A}_{\text{sub-lead}},
\end{equation}
where $\Delta_{Xg}=(p_X+q)^2-\mu_X^2$ and ${\cal A}_{P+g}$ (${\cal
  A}_{D+g}$) is the amplitude for on-shell production (decay) of the
unstable particle with an additional real gluon. The first two terms
in \Eqn{eq:A_real} contribute to the leading resonant real-emission
amplitude, whereas ${\cal A}_{\text{sub-lead}}$ contains terms
sub-leading in $\delta$.  Note that in this picture there is no 
topology with gluon emission from the unstable particle. In the corresponding Feynman
diagram either of the two unstable-particle propagators can be
resonant. In fact, when the emitted gluon is soft, $q \sim m_X
\delta$, both propagators can simultaneously be resonant,
$p_X^2-m_{0,X}^2 \sim (p_X+q)^2-m_{0,X}^2 \sim m_X^2 \delta$. However,
it can be shown~\cite{Campbell:2004ch} that the product of two
unstable-particle poles can be written as the sum of two terms
containing a single massive pole and a soft singularity
\begin{equation}
\frac{1}{(p_X^2-m_{0,X}^2) ((p_X+q)^2-m_{0,X}^2)}
=\frac{1}{2 p_X \cdot q} \left[\frac{1}{p_X^2-m_{0,X}^2}
- \frac{1}{(p_X+q)^2-m_{0,X}^2}\right] \, ,
\end{equation}
so that the parameterization \Eqn{eq:A_real} still holds true.

Following \Eqn{eq:A_real}, the squared amplitude, ${\cal
  M}_\text{real}\equiv |{\cal A}_\text{real}|^2$, can be written as
\begin{equation} \label{eq:M_leading} 
{\cal M}_\text{real} = 
\frac{|{\cal A}_{P+g} {\cal A}_D|^2}{|\Delta_X|^2}
+\frac{|{\cal A}_{P} {\cal A}_{D+g}|^2}{|\Delta_{Xg}|^2} 
+2 \text{Re}\left[\frac{({\cal A}_{P+g} {\cal A}_{D}) 
  ({\cal A}_P {\cal A}_{D+g})^\ast}{\Delta_X \Delta_{Xg}^\ast}\right]
+ \ldots
\end{equation}
The first two terms can be interpreted as factorizable real
corrections to either production or decay of the unstable particle
$X$, while the interference term gives rise to non-factorizable
corrections. Note that the interference term gives a sizeable
contribution only if both $\Delta_X$ and $\Delta_{Xg}$ are resonant,
i.e. when the emitted gluon is soft. This shows that, for the real
matrix element as well, non factorizable and off-shell effects are
encoded by soft radiation.  The omitted terms in \Eqn{eq:M_leading},
denoted by the ellipses, are suppressed by $\alpha_s \delta \sim
\delta^{3/2}$ compared to the leading tree-level contribution and
first contribute to the matrix element at N$^3$LO in the
power-counting.  Hence, they can be neglected in most calculations
relevant for hadron-collider phenomenology. However, we point out that
they can, in principle, be computed to the desired accuracy in
$\delta$, in a way similar to the expansion of the Born amplitude,
\Eqn{eq:ampl_tree_exp}.

\subsection{Cross sections}
\label{sec:EFT_cross-section}

The real and the virtual amplitudes squared have been split into a
hard factorizable part and a soft non-factorizable part. The hard part
is further divided into corrections to the production and
corrections to the decay. As usual, the virtual and real cross
sections, obtained by integrating the squared matrix elements over
phase space, have infrared and/or collinear singularities.

The cancellation of these singularities between real and virtual
corrections takes place separately for all parts of the cross
section. The (hard) production part typically has initial-state
collinear singularities that have to be absorbed into the parton
distribution functions. It can additionally develop infrared singularities 
and, when particles other than the unstable particle are present in the 
final state of the production subprocess, can develop final-state collinear 
singularities as well. 
These singularities cancel against the corresponding real singularities of
the production part. The (hard) decay part typically has final state
collinear singularities and infrared singularities that cancel against
the corresponding real singularities.

Finally, the non-factorizable soft corrections only have infrared, but
no collinear singularities. Once again, these cancel between the real and
virtual corrections. Also, with this cancellation, all explicit scale
dependence vanishes and the non-factorizable corrections only depend
implicitly on the renormalization and factorization scale. Because the
soft emission is governed by a soft scale, $\mu_s\sim m_X \delta$, it
is natural to choose a scale of this order for the coupling from the
additional soft emission. The factorization scale and the scale
related to the born term are not affected by this and should be kept
at the usual hard scale.

\subsection{Mass scheme}
\label{sec:EFT_mass-scheme}

The results given in this section have been expressed in terms of the
unstable-particle mass in a generic renormalization scheme,
$\hat{m}_X$, which is related to the gauge-invariant complex pole of
the propagator by $\mu_X^2=\hat{m}_X^2+\hat{m}_X \Omega_X$. While the
EFT approach is not limited to a particular scheme, it is more
naturally formulated in a class of schemes related to the pole scheme
by $(\hat{m}_X-m_X)/m_X \sim \delta$, which have the property of
preserving the EFT scaling $p_X^2-\hat{m}_X^2 \sim m_X^2 \delta$ under
renormalization. The $\overline{\text{MS}}$ is not such a scheme since
$(\hat{m}_{X,\overline{\text{MS}}}-m_X)/m_{X,\overline{\text{MS}}}
\sim \alpha_s \sim \sqrt{\delta}$ and the finite-width expansion
becomes less transparent in this case. The pole mass on the other
hand, which is the natural mass choice for unstable electroweak gauge
bosons and other particles which do not couple to strong interactions,
is affected by long-distance ambiguities related to QCD renormalons
\cite{Bigi:1994em, Beneke:1994sw,Smith:1996xz} in the case of heavy
quarks. This makes a precise definition and extraction of this
parameter problematic. It is possible to circumvent these issues with
the use of so-called ``threshold masses" \cite{Bigi:1994em,
  Beneke:1998rk,Hoang:1999zc,Pineda:2001zq} or ``jet masses"
\cite{Fleming:2007qr,Fleming:2007xt}, which preserve the EFT counting
and are not affected by renormalon ambiguities. The implication of the
mass-renormalization scheme choice in the case of $t \bar{t}$
production is discussed in Section~\ref{sec:mass}.

\section{Top-antitop production at hadron colliders}
\label{sec:amplitudes}

At hadron colliders the main top-production mechanism is the
production of a $t \bar{t}$ pair, which proceeds through strong
interactions\footnote{While the electroweak-mediated production of a
  single top is numerically only 3-4 times smaller than top-pair
  production, a much larger background makes the extraction of the
  signal in this channel more challenging.}. At Born level two
partonic processes contribute to the cross section: $q \bar{q}
\rightarrow t \bar{t}$, which dominates at Tevatron, and $gg
\rightarrow t \bar{t}$, which is the dominant production channel at
the LHC. Beyond leading order other partonic production channels
contribute as well. Next-to-leading order QCD corrections to the
tree-level results, both at the inclusive and differential level, have
been known for more than 20 years \cite{Nason:1987xz,
  Nason:1989zy,Beenakker:1990maa,Mangano:1991jk} and have been
recomputed recently for helicity-specific amplitudes using
generalized-unitarity methods \cite{Badger:2011yu}. One-loop
electroweak contributions are also known \cite{Beenakker:1993yr,
  Moretti:2006nf,Kuhn:2006vh,Bernreuther:2008aw,Hollik:2007sw}.  At
next-to-next-to-leading order in QCD several ingredients and
partial results have been known for some time
\cite{Czakon:2007ej,Czakon:2007wk,Dittmaier:2007wz,Dittmaier:2008uj,Czakon:2008zk,Anastasiou:2008vd,Bonciani:2008az,Bonciani:2009nb,Melnikov:2010iu,Czakon:2010td,Bonciani:2010mn,Bierenbaum:2011gg}
and the full NNLO calculations for the fermion-initiated \cite{Baernreuther:2012ws,Czakon:2012zr}
and $g q(\bar{q})$ \cite{Czakon:2012pz} production channels have been recently completed. 
In addition, resummation of Coulomb singularities and soft logarithms
up to next-to-next-to-leading logarithmic accuracy has been studied by
several groups \cite{Kidonakis:2010dk,Ahrens:2011mw,Cacciari:2011hy,Beneke:2011mq,Beneke:2012eb}.

All the aforementioned results were computed in the approximation of stable
tops. Results in the NWA, including the on-shell decay
of the top-antitop pair to the physical final state $W^+W^{-}
b\bar{b}$, were given in
Refs.~\cite{Bernreuther:2004jv,Melnikov:2009dn,Bernreuther:2010ny,Campbell:2012uf}.
The fully differential semi-leptonic decay of on-shell top quarks is also now known to 
NNLO \cite{Gao:2012ja,Brucherseifer:2013iv}.
Recently the full NLO QCD calculation of the off-shell top-pair production and
decay process in the complex-mass scheme was performed by two
separate groups \cite{Denner:2010jp, Bevilacqua:2010qb,Denner:2012yc}\footnote{The
  calculation of Ref.~\cite{Bevilacqua:2010qb} treats also the leptonic $W$
  decays completely off-shell, while in Ref.~\cite{Denner:2010jp} the $W$
  decays are described in narrow-width approximation. The effects of
  finite $W$-width effects in $t\bar{t}$ production have been
  investigated recently in Ref.~\cite{Denner:2012yc}.}.  As expected,
finite-width effects were found to be small for the total cross
section. However, larger effects of up to tens of per cent were
observed for more exclusive kinematical distributions
\cite{AlcarazMaestre:2012vp}, thus confirming the necessity of a careful
treatment of finite-width effects both in $t \bar{t}$ production, and more
generally, in processes involving unstable particles.

In this section we present the calculation of the LO and NLO virtual
and real amplitudes for $t \bar{t}$ production in the EFT framework,
as a non-trivial example of application of the formalism described in
Section \ref{sec:EFT}.  We will focus on the $q \bar{q}$ production
channel,
\begin{equation}
q \bar{q} \rightarrow t \bar{t} \rightarrow W^+ W^- b \bar{b} \nonumber\\
\end{equation}
where the decay of the two $W$ bosons to leptons is understood. 
Our notation and conventions for momentum and colour
labelling is given in Section~\ref{sec:tree} where we also give
explicit expressions for the leading tree-level helicity amplitudes.
The calculation of soft and hard virtual corrections as well as the
renormalization necessary at NLO are detailed in
Section~\ref{sec:virtual}, while the implementation of real
corrections is described in Section~\ref{sec:real}.
  
\subsection{Leading tree-level amplitude}
\label{sec:tree}

Throughout this work we adopt the following momentum labelling
\begin{equation} \label{eq:proc_tree}
q(p_1; c_1) \bar{q}(p_2; c_2) 
\rightarrow b \bar{b} W^+ W^-
\rightarrow b(p_3; c_3) \bar{b}(p_4; c_4) 
\bar{l}_1(p_5) \nu_1 (p_6) l_2 (p_7) \bar{\nu}_2 (p_8) \, ,
\end{equation}
with $c_1 ... c_4$ the colour indices of the four external quarks. The
leptonic decay of the two $W$-bosons is treated in the NWA. 
For sake of simplicity, we introduce the shortcuts
\begin{eqnarray}
p_t &=& p_3+p_5+p_6 \, ,\nonumber \\
p_{\bar{t}} &=& p_4+p_7+p_8 \, ,\nonumber
\end{eqnarray}
and the on-shell projections $\bar{p}_t$ and $\bar{p}_{\bar{t}}$, with
$\bar{p}_t^2=\bar{p}_{\bar{t}}^2=\mu_t^2$, which are
necessary\footnote{Note that in an NLO calculation one can set
  $\bar{p}_t^2=\bar{p}_{\bar{t}}^2=m_t^2$ instead, since the term $-i
  m_t \Gamma_t \sim m_t^2 \delta$ is formally an NNLO
  contribution.} for the systematic expansion of the amplitudes in
$\delta$.  As usual $\mu_t^2$ denotes the complex pole of the
propagator, $\mu_t^2=m_t^2-i m_t \Gamma_t$.  We also introduce the
abbreviations
\begin{equation} \label{eq:Deltat}
\Delta_t \equiv p_t^2-\mu_t^2 \hspace{1cm}
\Delta_{\bar{t}} \equiv p_{\bar{t}}^2-\mu_t^2  \, ,
\end{equation}  
for the resummed inverse top and antitop propagator. The
expression~(\ref{eq:Deltat}) is valid in the pole scheme, i.e $m_t$
here denotes the pole mass. The conversion to other renormalization
schemes is discussed in Section~\ref{sec:mass}.  The amplitudes for
the $t \bar{t}$ production process are decomposed onto a basis of
colour-state operators, given by
\begin{equation}\label{eq:col}
{\cal O}^{(1)}_{\{c\}}=\frac{1}{N_c}\delta_{c_1 c_2} \delta_{c_3 c_4} \hspace{2 cm}
{\cal O}^{(2)}_{\{c\}}=\frac{2}{\sqrt{N_c^2-1}}T^A_{c_2 c_1} T^A_{c_3 c_4} \, ,
\end{equation}
with $T^A$ denoting the generators of the $SU(3)$ algebra in the fundamental
representation and $N_c=3$. The two operators describe the $s$-channel
exchange of a colour-singlet and colour-octet state respectively, and
satisfy the orthonormality condition
\begin{equation}
\sum_{\{c\}} {\cal O}^{(i)}_{\{c\}} {\cal O}^{(j)\ast}_{\{c\}}=\delta_{i j} \, .
\end{equation}

\begin{figure}[t!]
\begin{center}
\includegraphics[width=0.7 \linewidth]{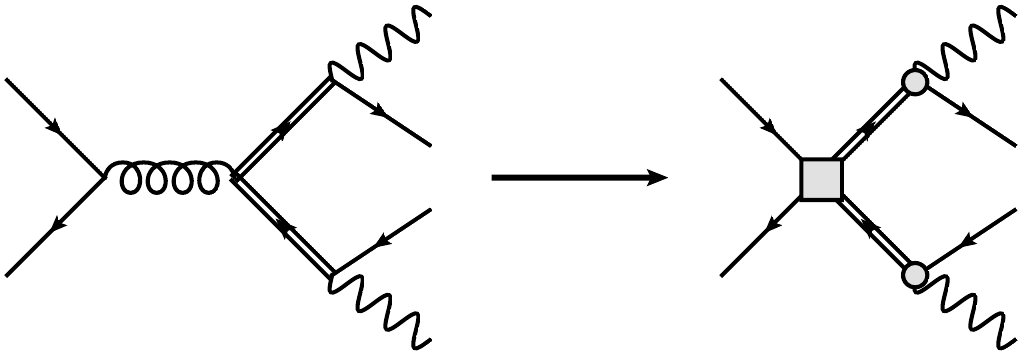}
\end{center}
\caption{Leading tree-level doubly-resonant contribution to $q \bar{q}
  \rightarrow b \bar{b} W^+ W^-$ in the SM and its EFT
  counterpart. The grey square and circles represent EFT production
  and decay matching coefficients.}
\label{fig:tree}
\end{figure}

At the Born level both doubly-resonant Feynman diagrams containing
an intermediate top and antitop pair, as well as singly-resonant and
non-resonant diagrams, contribute to the amplitude for the
process~(\ref{eq:proc_tree}).  The dominant tree-level
contribution to the amplitude is given by the leading term in the expansion 
in $\delta$ of the left Feynman diagram in
Figure~\ref{fig:tree}. In the effective-theory framework, the
interaction mediated by the intermediate $s$-channel gluon is replaced
by a four-fermion contact interaction (right diagram in
Figure~\ref{fig:tree}) of the form
\begin{equation}
\frac{i g_s^2}{\hat{s}} \gamma^\mu_{\alpha_2 \alpha_1} 
\gamma_{\mu,\alpha \bar{\alpha}} T^A_{c_2 c_1} T^A_{c \bar{c}} \, ,
\end{equation} 
with $\hat{s}$ the partonic centre-of-mass energy, $\alpha_1,\,
\alpha_2,\, \alpha,\, \bar{\alpha}$ the Lorentz indices of the initial
quark and antiquark and final top and antitop, and $c_1,\, c_2,\, c,
\, \bar{c}$ the corresponding colour indices.  The tree level
amplitude reads (omitting the on-shell decay of the two $W$s)
\begin{equation} \label{eq:tree_LO}
{\cal A}_\text{tree}=\frac{\sqrt{N_c^2-1}}{2} {\cal O}^{(2)}_{\{c\}} 
{\cal A}^{(0)} \,  ,
\end{equation} 
with
\begin{equation} \label{eq:A_LO}
{\cal A}^{(0)} =
\frac{g_s^2 g^2}{2 \hat{s} \Delta_t \Delta_{\bar{t}}} 
\bar{v}(\bar{p}_2) \gamma^\mu u(\bar{p}_1) \bar{u}(\bar{p}_3) 
\dirac{\epsilon}^\ast_+(\dirac{\bar{p}}_t+m_t) \gamma_\mu
 (-\dirac{\bar{p}}_{\bar{t}}+m_t) \dirac{\epsilon}^\ast_- v(\bar{p}_4) \, .
\end{equation}
$\epsilon_{\pm}$ represent the polarization vectors of $W^{\pm}$.
Note that the numerator of ${\cal A}^{(0)}$ depends on the projected
on-shell momenta which guarantees the gauge-invariance of the
result. The contribution of resonant and non-resonant diagrams will be
discussed in Section~\ref{sec:NR}.

Using \Eqn{eq:counting} one can easily determine the scaling of the
leading amplitude, which is
\begin{equation}
{\cal A}^{(0)} \sim 
\frac{\alpha_s \alpha}{\Delta_t \Delta_{\bar{t}}} \sim \delta^{-1/2} \, .
\end{equation}
As a consequence of the relative scaling of the top-quark width and
the strong coupling constant, $\alpha_s \sim \sqrt{\Gamma_t/m_t}$, the
expansion of the amplitude is organized in powers of
$\delta^{1/2}$. Thus in the following we define as N$^k$LO all
contributions to the amplitude that scale as $\delta^{(k-1)/2}$. In
particular, NLO corrections, which we will focus on in this paper,
scale as $\delta^0$. Potentially, these include sub-leading tree-level
terms, ${\cal O}(\alpha_s)$ hard and soft virtual corrections to the
leading doubly-resonant amplitude \Eqn{eq:tree_LO}, and real QCD
corrections.  These three contributions will be discussed in turn in
the following sections.
  
\subsection{Sub-leading tree-level contributions}

\label{sec:NR}
\begin{figure}[t!]
\begin{center}
\includegraphics[width=0.8 \linewidth]{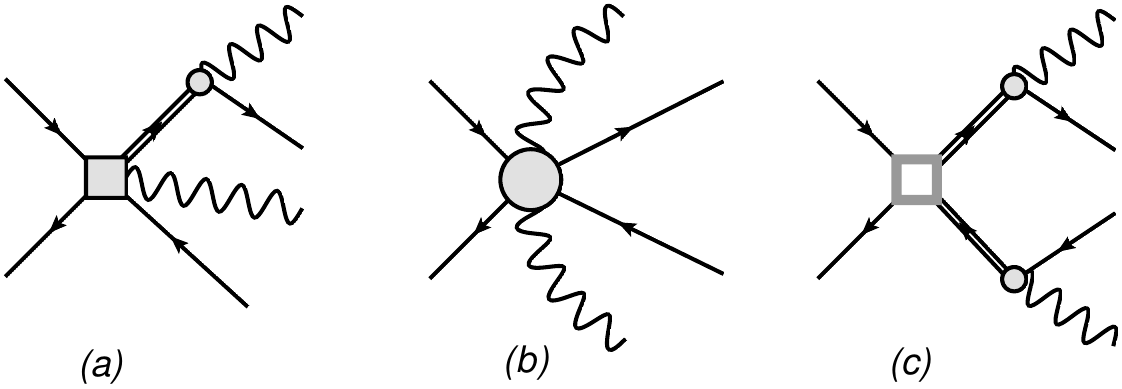}
\end{center}
\caption{Subleading tree-level diagrams in the effective theory: (a)
  singly-resonant diagrams, (b) non-resonant diagrams, (c)
  doubly-resonant diagrams with electroweak matching coefficients
  (represented by the grey and white square in the figure). See the
  text for explanation.}
\label{fig:NR}
\end{figure}
The first class of corrections which could potentially contribute at
NLO in $\delta$ is represented by terms arising from sub-leading
tree-level diagrams, which are shown in Figure~\ref{fig:NR}. These
include QCD singly-resonant and non-resonant topologies (diagrams (a)
and (b)). Since the effective-theory counting of \Eqn{eq:counting}
relates the expansion in $\alpha_s$ to the one in the electroweak
coupling constant, electroweak diagrams must also be taken into
account beyond LO in $\delta$. For all practical purposes it is
sufficient to consider the leading doubly-resonant topologies
corresponding to diagram (c) in Figure~\ref{fig:NR}. Here we use
scaling arguments to determine the size of these terms compared to the
leading-order amplitude, \Eqn{eq:A_LO}.
  
Contributions with the structure of diagram (a) correspond to the
terms on the second line of \Eqn{eq:ampl_tree_exp}, where one of
the resonant propagators is either cancelled by higher-order contributions
originating from the expansion around the pole, or is not present in the
first place. In the effective theory they are described by
five-particle contact interactions, whose matching coefficients are
proportional to $g_s^2 g_{ew}$. Their contribution to the amplitude thus
scales as
\begin{equation}
\frac{\alpha_s \alpha_{ew}}{\Delta_t} \sim \delta^{1/2} \, ,
\end{equation}
which makes them an NNLO correction. Their interference with the
leading Born amplitude gives a NNLO correction to the cross section,
while the square of diagram (a) contributes to the cross section initially 
at N$^4$LO.  Non-resonant QCD topologies correspond to six-particle
effective interactions (diagram (b)) with matching coefficients
starting at order $g_s^2 g_{ew}^2$. Since they do not contain intermediate
resonant propagators, these scale as
\begin{equation}
\alpha_s \alpha_{ew} \sim \delta^{3/2}
\end{equation}
and contribute to the cross section as N$^4$LO in $\delta$,
being suppressed by $\delta^2$ compared to the leading Born amplitude.

The doubly resonant electroweak diagrams (diagram (c)) scale as 
\begin{equation}
\frac{\alpha_{ew}^2}{\Delta_t \Delta_{\bar{t}}} \sim \delta^0 \, .
\end{equation} 
Note that in this case the suppression is given by the electroweak
nature of the matching coefficients ($\sim g_{ew}^2$) of the four-particle
production operator, rather than by missing resonant propagators. The
interference of diagram (c) with the LO amplitude thus gives an NLO
correction. However, due to the colour structure of the electroweak
production operators, this interference is non-vanishing only for a $b
\bar{b}$ initial state. This contribution is numerically suppressed due to the
smallness of the bottom-quark PDF inside the proton and is in
practice negligible ($<0.1 \%$ of the leading doubly-resonant
QCD-mediated diagram). The squared electroweak matrix element counts
as an NNLO correction and in this case there is generally no
suppression arising from the parton luminosities, since operators with
any light-flavour quark in the initial state contribute.

To conclude, if one neglects the (accidentally) tiny NLO contribution
from the interference of diagram (c) with ${\cal A}^{(0)}$, the first
Born-level corrections to \Eqn{eq:A_LO} arise at NNLO from the
interference of singly-resonant QCD diagrams with the leading
amplitude and from the squared matrix element of electroweak
doubly-resonant diagrams. Being suppressed by $\delta \sim
\Gamma_t/m_t$ compared to \Eqn{eq:A_LO} they are expected to
contribute to the cross section at the percent level. This is
investigated more quantitatively in Section \ref{sec:res}, in
particular to test whether the EFT counting applies in practice to the
case of $t \bar{t}$ production. As we will see, sub-leading terms
account for at most a few percent of the cross section which is in
good agreement with the expectation from the EFT scaling arguments.

\subsection{Virtual QCD corrections}
\label{sec:virtual}

\begin{figure}[t!]
\begin{center}
\includegraphics[width=0.9 \linewidth]{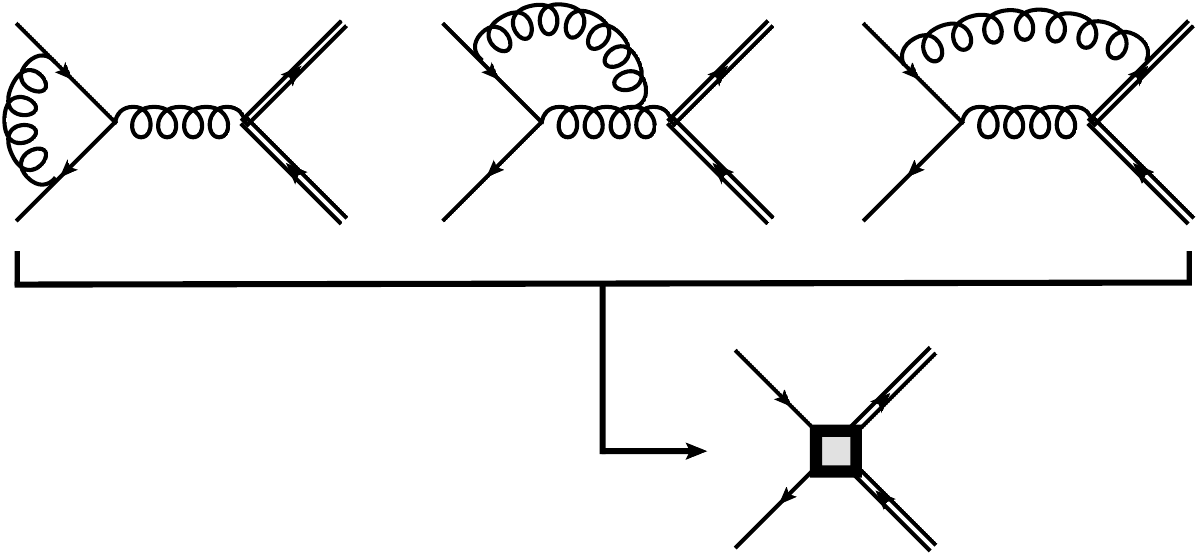}
\end{center}
\caption{A sample of one-loop diagrams (above) contributing to the
  ${\cal O}(\alpha_s)$ corrections to the production matching
  coefficient (below).}
\label{fig:tT_C1}
\end{figure}
\begin{figure}[t!]
\begin{center}
\includegraphics[width=0.7 \linewidth]{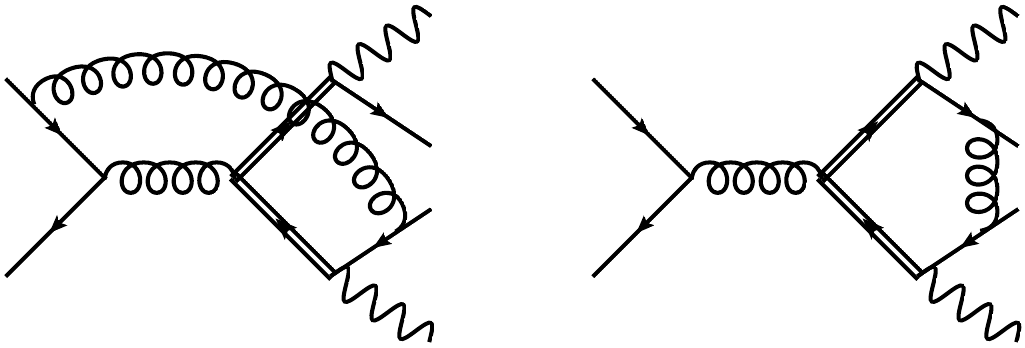}
\end{center}
\caption{Two examples of one-loop diagrams which do not contribute to
  the hard matching coefficients at NLO in $\delta$. See the text for
  further explanation.}
\label{fig:hard_sub}
\end{figure}

As explained in Section~\ref{sec:EFT_virtual}, virtual corrections in
the EFT approach are classified into corrections to the hard matching
coefficients of the production and decay effective vertices as well as soft
corrections to the EFT matrix elements. The former encode physics at
the large scale, $q\sim m_t$, whilst the latter describe long-scale
physics, $q\sim m_t \delta$.  To obtain NLO accuracy, $\alpha_s$
corrections to the hard matching coefficients and one-loop soft
corrections are required. As pointed out in
Section~\ref{sec:EFT_virtual}, ${\cal O}(\alpha_s)$ hard matching
coefficients can be related to one-loop corrections to the amplitudes
for production and decay of on-shell tops. For the production part we
use the results of Ref.~\cite{Badger:2011yu}, whereas the results for
the decay can for example be found in Ref.~\cite{Campbell:2004ch}.  A
sample of the diagrams contributing to the production matching
coefficient is shown in Figure~\ref{fig:tT_C1}.  As for the toy model
discussed in Section~\ref{sec:EFT_virtual}, hard loops connecting
initial and final-state particles, or final-state particles
originating from different decays, are suppressed by higher powers of
$\delta$. Consider, for example, the two diagrams shown in Figure
\ref{fig:hard_sub}. If the momentum flowing in the loop is hard, the
intermediate antitop propagator in the first diagram is
far off-shell. Thus the diagram is suppressed by $\alpha_s
\Delta_{\bar{t}} \sim \delta^{3/2}$ compared to the leading amplitude
\Eqn{eq:A_LO}, which makes it a N$^3$LO correction. The pentagon
diagram, in which both top and antitop propagators are off shell, is
even more suppressed, $\sim \alpha_s \Delta_t \Delta_{\bar{t}} \sim
\delta^{5/2}$, and contributes to the amplitude starting at
N$^5$LO. These contributions do not have to be included in a
calculation that aims for NLO accuracy in $\delta$, thus dramatically
reducing the computational complexity of the EFT calculation compared
to that in the complex-mass scheme.

\begin{figure}[t!]
\begin{center}
\includegraphics[width=0.95 \linewidth]{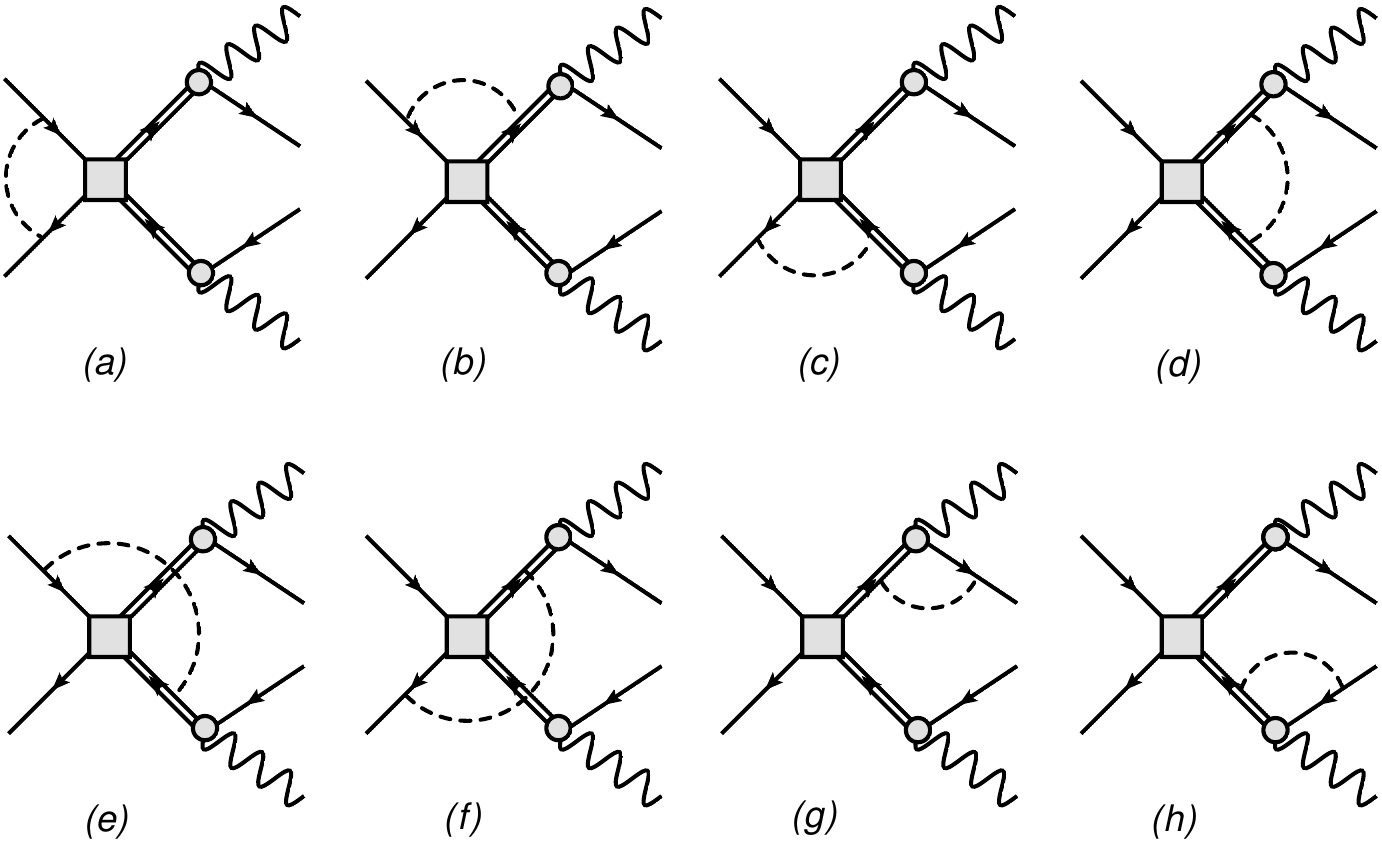}
\end{center}
\caption{Soft triangle diagrams contributing at NLO in the EFT counting. Dashed lines represent soft gluons.}
\label{fig:triangles}
\end{figure}

\begin{figure}[t!]
\begin{center}
\includegraphics[width=0.95 \linewidth]{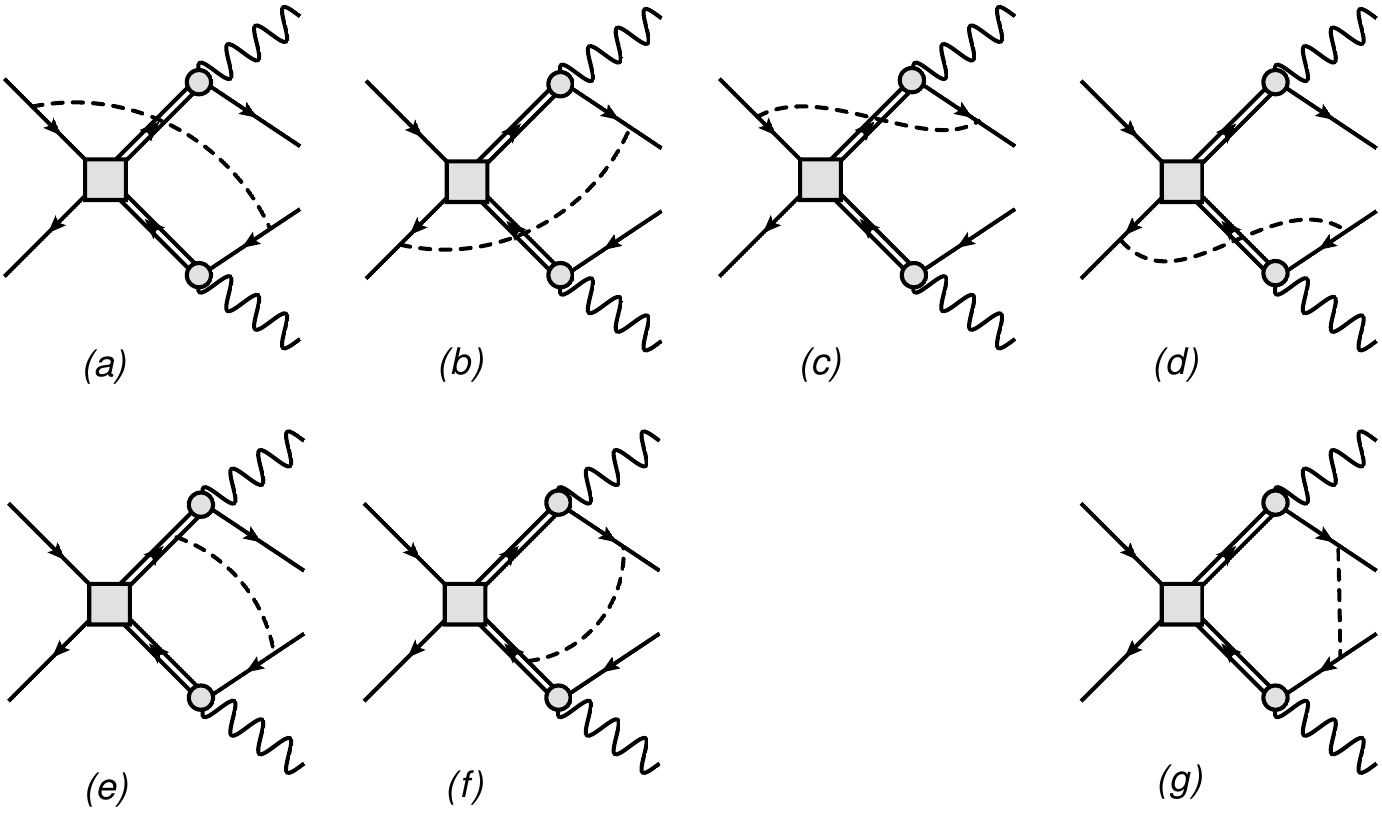}
\end{center}
\caption{Soft box and pentagon diagrams contributing at NLO in the EFT
  counting. Dashed lines represent soft gluons.} 
\label{fig:boxes}
\end{figure}

Contrary to the hard corrections, non-factorizable soft diagrams that
connect initial and final states do contribute already at NLO in
$\delta$, since the two intermediate top propagators remain resonant
if the loop momentum is soft. The complete set of soft one-loop
diagrams contributing to the amplitude at order $\delta^0$ consists of:
\begin{itemize}
\item 2 self-energy corrections, to the top and antitop resonant propagator,
\item 8 triangle diagrams, shown in Figure \ref{fig:triangles},
\item 6 box diagrams, shown in Figure \ref{fig:boxes},
\item 1 pentagon diagram, also shown in Figure \ref{fig:boxes} (right bottom corner).
\end{itemize}
Note that the renormalization of the bare couplings and masses only affects
the hard matching coefficients, but does not contribute to the
soft corrections. The contribution of the soft diagrams listed above
can be obtained by expanding full QCD diagrams in the soft region. 
Only a consistent expansion in $\delta$ guarantees the gauge 
invariance of the final result. Most of the
integrals shown in Figures~\ref{fig:triangles} and \ref{fig:boxes}
were already computed for the case of single-top production
\cite{Falgari:2010sf,Falgari:2011qa}, with the new soft integrals required
given by the triangle in Figure~\ref{fig:triangles}d, the boxes
\ref{fig:boxes}e and \ref{fig:boxes}f, and the pentagon. As a further
example of how the method of regions works we will now briefly
discuss the calculation of the triangle diagram~\ref{fig:triangles}d.

All soft integrals can be written in terms of the colour operators
\Eqn{eq:col}, the colour-stripped leading-order amplitude
\Eqn{eq:A_LO} and scalar functions depending on the invariants $s_{ij}
\equiv 2 p_i \cdot p_j$ built from the external momenta. In
particular, the contribution to the amplitude of the triangle in
Figure~\ref{fig:triangles}d can be written as
\begin{equation}
{\cal A}^{(1),V}_{3,d} = 
-\frac{\sqrt{N_c^2-1}}{4 N_c} {\cal O}^{(2)}_{\{a\}} 
{\cal A}^{(0)} I_{t \bar{t}} \, ,
\end{equation}
with the scalar integral $I_{t \bar{t}}$ defined as 
\begin{equation}\label{eq:Ittbar}
I_{t \bar{t}}=16 \pi i \alpha_s (\bar{p}_t \cdot \bar{p}_{\bar{t}}) 
\frac{e^{\epsilon \gamma_E} \mu^{2 \epsilon}}{(4 \pi)^\epsilon} \int
\frac{d^d q}{(2 \pi)^d} \frac{1}{q^2} 
\frac{1}{\Delta_t-2 \bar{p}_t \cdot q}
\frac{1}{\Delta_{\bar{t}}+2 \bar{p}_{\bar{t}} \cdot q}
\end{equation}
and $d=4-2 \epsilon$.  The simple expression of $I_{t \bar{t}}$
follows from the eikonal form of the QCD fermion-gluon interaction
vertex in the soft limit, $\bar{u}(p) \gamma^\mu (\slash{p}_t\pm
\slash{q}+m_t) \sim 2 \bar{p}_t^\mu \bar{u}(p)$, and from the
expansion of the propagators in the integrand of the full QCD triangle
integral,
\begin{equation}
(p_t \pm q)^2-\mu_t^2 = 
p_t^2 \pm 2 p_t \cdot q+q^2-\mu_t^2 
\sim \Delta_t \pm 2 \bar{p}_t \cdot q \, , 
\end{equation} 
where the quadratic term $q^2 \sim \delta^2$ must be dropped, being
sub-leading compared to the term $ \Delta_t \pm 2 \bar{p}_t \cdot q
\sim \delta$. Note that in \Eqn{eq:Ittbar} the off-shell top and
antitop momenta $p_t$ and $\bar{p}_t$ have been replaced by the
on-shell projections $\bar{p}_t$ and $\bar{p}_{\bar{t}}$, except in
the combinations $p_t^2-\mu_t^2 \equiv \Delta_t$ and
$\bar{p}_t^2-\mu_t^2 \equiv \Delta_{\bar{t}}$. In addition, finite-width
effects have been resummed into the top propagators, in accordance
with the EFT Lagrangian \Eqn{eq:kinET}.
 
Using Feynman parameters and standard integration techniques
\Eqn{eq:Ittbar} can be reduced to the one-dimensional integral
\begin{equation}
I_{t \bar{t}} = 
\frac{\alpha_s}{2 \pi} \eta e^{\epsilon \gamma_E} 
\Gamma(1-\epsilon) \Gamma(2 \epsilon)
\int_0^\infty dy 
\left[(y-\xi_+) (y-\xi_-)\right]^{-1+\epsilon} 
 \left(\frac{-\Delta_t-y \Delta_{\bar{t}}}{m_t \mu}\right)^{-2 \epsilon} \, ,
\end{equation}
where we have introduced the invariant $\eta=2 \bar{p}_t \cdot
\bar{p}_{\bar{t}}/m_t^2$ and  
\begin{equation}\label{eq:xi}
\xi_{\pm} = \frac{\eta \pm \sqrt{\eta^2-4+i o_+}}{2} \, ,
\end{equation}
with $i o_+$ an infinitesimal and positive imaginary part.
Expanding $I_{t \bar{t}} $ in $\epsilon$ one obtains
\begin{equation} \label{eq:Itt_final}
I_{t \bar{t}}  = \frac{\alpha_s}{4 \pi} \eta 
\left\{I_{t \bar{t}}^{(-1)} 
\left[\frac{1}{\epsilon}-2 \ln \left(-\frac{\Delta_t}{m_t \mu} \right)\right]
+I_{t \bar{t}}^{(0)}\right\} \, ,
\end{equation}
where the integrals $I_{t \bar{t}}^{(-1)}$ and $I_{t \bar{t}}^{(0)}$ are given by
\begin{equation}
I_{t \bar{t}}^{(-1)} = -\frac{\ln(-\xi_+)-\ln(-\xi_-)}{\xi_+-\xi_-}\, ,
\end{equation}
\begin{eqnarray} \label{eq:Itt0}
I_{t \bar{t}}^{(0)} 
&=& \frac{1}{\xi_+ -\xi_-} \left\{- \frac{1}{2} \ln^2(-\xi_+) + \frac{1}{2} \ln^2(-\xi_-)^2  
-\ln(-\xi_+) \ln(\xi_+-\xi_-)\right.\nonumber\\
 &&\left. +\ln(-\xi_-) \ln(\xi_--\xi_+)+\frac{1}{2} \ln^2(\xi_+-\xi_-)-\frac{1}{2} \ln^2(\xi_--\xi_+) \right.\nonumber\\
 &&\left.+\text{Li}_2\left(\frac{\xi_+}{\xi_+ - \xi_-}\right)-\text{Li}_2\left(\frac{\xi_-}{\xi_- - \xi_+}\right) - 
 \ln^2\left(\frac{\Delta_{\bar{t}}}{\Delta_t + \Delta_{\bar{t}} \xi_+}\right) +\ln^2\left(\frac{\Delta_{\bar{t}}}{\Delta_t + \Delta_{\bar{t}} \xi_-}\right) \right.\nonumber\\
 &&- 
 2 \ln(-\xi_+) \ln \left(\frac{\Delta_t}{\Delta_t+\Delta_{\bar{t}} \xi_+}\right) + 
 2 \ln(-\xi_-) \ln \left(\frac{\Delta_t}{\Delta_t+\Delta_{\bar{t}} \xi_-}\right)\nonumber\\
     &&\left.- 
 2 \text{Li}_2\left(\frac{\Delta_{\bar{t}} \xi_+}{\Delta_t + \Delta_{\bar{t}} \xi_+}\right) +2 \text{Li}_2\left(\frac{\Delta_{\bar{t}} \xi_-}{\Delta_t + \Delta_{\bar{t}} \xi_-}\right)
\right\} \, .
\end{eqnarray}
Note that the only explicit scale dependence in \Eqn{eq:Itt_final} is
due to the term $\ln \left(-\Delta_t/(m_t \mu) \right)$, indicating 
that the natural scale choice for non-factorizable corrections is
$\mu_{\text{soft}}  \sim \Delta_t/m_t \sim \Gamma_t$.  All other terms in
\Eqn{eq:Itt_final} have no explicit $\mu$ dependence and do not
contain potentially large logarithms. This is a common feature of all
soft contributions in Figures~\ref{fig:triangles} and \ref{fig:boxes}.
The calculation of the missing box and pentagon integrals closely follows
the one outlined above for the triangle and will not be
given explicitly here. 

\subsection{Real QCD corrections}
\label{sec:real}

\begin{figure}[t!]
\begin{center}
\includegraphics[width=0.8 \linewidth]{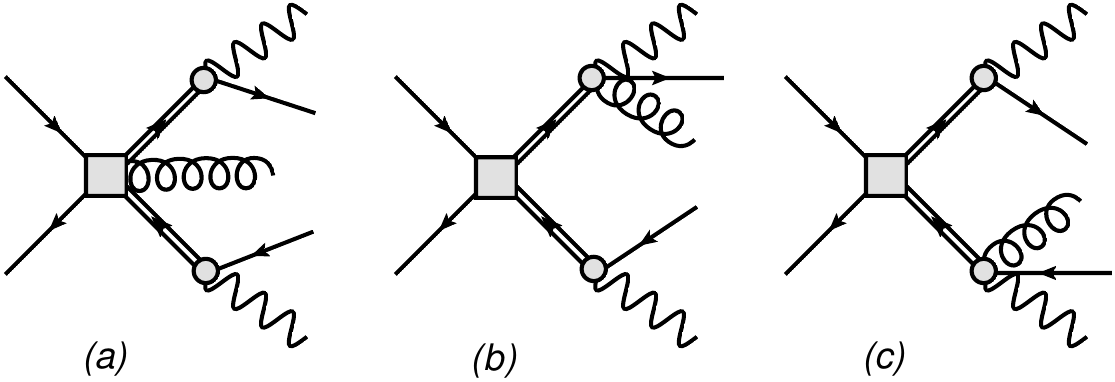}
\end{center}
\caption{Real-corrections diagrams contributing to the cross section
  at NLO in $\delta$} 
\label{fig:real}
\end{figure}

At NLO real QCD diagrams must also be taken into account. As discussed
in Section~\ref{sec:EFT_real} and illustrated in Figure~\ref{fig:real}, 
any real emission diagram can be split
into a linear combination of terms containing a single resonant
propagator. In turn, these can be interpreted as corrections to either the
production or decay subprocesses and  
their contribution to the amplitude is obtained
from the expansion of the full QCD matrix elements around on-shell
configurations. 
These
on-shell configurations are defined by $p_t^2=p_{\bar{t}}^2=\mu_t^2$
for diagram (a), $(p_t+k)^2=p_{\bar{t}}^2=\mu_t^2$ for diagram (b) and
$p_t^2=(p_{\bar{t}}+k)^2=\mu_t^2$ for diagram (c), with $k$ being the
momentum of the emitted gluon.
Note that here only gluonic corrections to the tree-level topology are
considered, but a full calculation which also includes the gluon-initiated
production channel would require the calculation of diagrams with a $g q(\bar{q})$
initial state at NLO.
 
Soft and collinear singularities are treated in the standard way,
by adding a suitable subtraction term to the real matrix element and
subtracting the corresponding integrated function from the virtual
corrections, such that the cancellation of poles is manifest and real
corrections can be safely computed in four dimensions. In particular,
for the calculation of the results given in Section~\ref{sec:res}, two
independent numerical codes were used, one implementing the FKS
subtraction method \cite{Frixione:1995ms,Frederix:2009yq} and the other, 
the Catani-Seymour dipole method \cite{Catani:1996vz,Catani:2002hc}. As
explained in Section~\ref{sec:EFT_cross-section}, the cancellation of
virtual poles against real ones occurs separately for
factorizable and non-factorizable corrections. For factorizable
corrections this follows from the cancellation of infrared and
collinear singularities for the on-shell production and decay
subprocesses.  On the other hand, the (infrared) poles of
non-factorizable virtual corrections (i.e. the sum of all soft virtual
contributions) are given by
\begin{eqnarray}
d \sigma_{V, \text{non-fact.}} &=& \frac{\alpha_s}{6 \pi \epsilon} \left[ 
16 + 7 \ln\left(\frac{s_{13}}{m_t^2}\right)+ 7 \ln\left(\frac{s_{24}}{m_t^2}\right) + 2 \ln\left(\frac{s_{14}}{m_t^2}\right) + 2 \ln\left(\frac{s_{23}}{m_t^2}\right) \phantom{\frac{1}{1}}\right.\nonumber\\
&& - 
 \ln\left(\frac{s_{34}}{m_t^2}\right) - 7 \ln\left(\frac{s_{1t}}{m_t^2}\right)  - 7 \ln\left(\frac{s_{2\bar{t}}}{m_t^2}\right)- 
 2 \ln\left(\frac{s_{1\bar{t}}}{m_t^2}\right) - 2 \ln\left(\frac{s_{2t}}{m_t^2}\right)\nonumber\\
 &&\left. - 8 \ln\left(\frac{s_{3t}}{m_t^2}\right) - 8 \ln\left(\frac{s_{4\bar{t}}}{m_t^2}\right)+\frac{\eta [\ln(-\xi_+)-\ln(-\xi_-)]}{2 (\xi_+-\xi_-)} \right] d \sigma_{\text{Born}}\,,
\end{eqnarray}
with $s_{i j}=2 p_i \cdot p_j$. These can be shown to coincide with
the pole structure of the production-decay interference terms in
\Eqn{eq:M_leading}.

\subsection{Mass-scheme choice}
\label{sec:mass}

The results shown in the previous sections have been computed in the
so-called ``pole-mass scheme", in which the renormalized mass
parameter is chosen to coincide with the real part of the complex pole
of the top-quark propagator. This particular mass scheme is affected
by ambiguities related to QCD renormalons \cite{Bigi:1994em,
  Beneke:1994sw,Smith:1996xz}, making a determination of the top-quark
mass at an accuracy better than $\Lambda_{\text{QCD}}$ conceptually
impossible. Avoiding these problems requires the use of a mass scheme which
is not affected by long-distance ambiguities. An example of such a
mass is the $\overline{\text{MS}}$-renormalized mass,
$m_t^{\overline{\text{MS}}}$. On the other hand the
$\overline{\text{MS}}$ mass presents the unpleasant feature of not
preserving the effective-theory counting (or, in other words, it
differs from the pole mass by terms of order $\alpha_s$), thus
spoiling the EFT counting scheme. More suitable choices are
represented by the so-called ``threshold masses" \cite{Bigi:1994em,
  Beneke:1998rk,Hoang:1999zc,Pineda:2001zq}, introduced in the context
of heavy-quark threshold physics, and ``jet masses"
\cite{Fleming:2007qr,Fleming:2007xt} studied for boosted top-jet
events in $e^+ e^-$-collisions. These masses have the attractive feature of
not being sensitive to low-momentum physics and of differing only by
amounts of order $\sim m_t \alpha_s^2 \sim m_t \delta$ from the pole
mass, thus preserving the effective-theory counting. In the following
we will consider the ``potential-subtracted" (PS) mass definition
\cite{Beneke:1998rk} as an alternative suitable mass scheme,
illustrating how this fits into the effective-theory calculation.

The relation between the pole mass and the PS mass can be found in
Ref.~\cite{Beneke:1998rk}, and up to second order in $\alpha_s$ 
(enough for applications in this work) reads
\begin{equation} \label{eq:PSmass}
m_t=m_{t,\text{PS}}+\mu_{\text{PS}} \left[
\frac{\alpha_s}{2 \pi} \delta_1^{\text{PS}}
+\left(\frac{\alpha_s}{2 \pi} \right)^2 \delta_2^{\text{PS}}\right]\, ,
\end{equation} 
with 
\begin{eqnarray}
\delta_1^{\text{PS}} &=& 2 C_F \nonumber\\
\delta_2^{\text{PS}} &=& 
C_F \left[a_1-b_0 \left(\ln\left(\frac{\mu_{\text{PS}}^2}{\mu_R^2}\right)
-2\right) \right]
\end{eqnarray}
and $a_1=31/3-10 n_f/9$, $b_0=11-2 n_f/3$. The cutoff
$\mu_{\text{PS}}$ has to be chosen of order $\sim m_t \alpha_s$
to preserve the EFT counting. Substituting the pole mass in favour of
the PS mass in the (renormalized) resummed top propagator one obtains
\begin{eqnarray}
\frac{1}{p_t^2-m_{t,\text{PS}}^2+i m_{t,\text{PS}} \Gamma_t
-\frac{\alpha_s}{\pi} \delta_1^{\text{PS}} \mu_{\text{PS}} m_{t,\text{PS}} 
-\frac{\alpha_s^2}{2 \pi^2} 
\delta_2^{\text{PS}} \mu_{\text{PS}} m_{t,\text{PS}}+\ldots}
\end{eqnarray}
with the ellipses denoting terms that scale as $\delta^2$ or
higher. Note that, counting $\mu_{\text{PS}} \sim \sqrt{\delta}$, the
term $-\alpha_s \delta_1^{\text{PS}} \mu_{\text{PS}} m_{t,\text{PS}}
/\pi$ scales as the leading term $p_t^2-m_{t,\text{PS}}^2+i
m_{t,\text{PS}} \Gamma_t \sim \delta$, and has to be kept in the
denominator of the propagator. The term proportional to $\alpha_s^2$
is suppressed by an additional $\sqrt{\delta}$ and can be re-expanded
to fixed-order in the expansion parameters,
\begin{eqnarray}
\frac{1}{p_t^2-m_{t,\text{PS}}^2+i m_{t,\text{PS}} \Gamma_t
-\frac{\alpha_s}{\pi} \delta_1^{\text{PS}} \mu_{\text{PS}}
m_{t,\text{PS}}} \left\{ 1+\frac{\frac{\alpha_s^2}{2 \pi^2}  
\delta_2^{\text{PS}} \mu_{\text{PS}} m_{t,\text{PS}}
}{p_t^2-m_{t,\text{PS}}^2+i m_{t,\text{PS}} \Gamma_t 
-\frac{\alpha_s}{\pi} \delta_1^{\text{PS}} \mu_{\text{PS}} m_{t,\text{PS}}}\right\} . 
\end{eqnarray}
Note that the replacement $m_t \rightarrow
m_{t,\text{PS}}+\frac{\alpha_s}{2 \pi} \delta_1^{\text{PS}}
\mu_{\text{PS}}+...$ in the numerator of the matrix elements does not
generate extra terms at NLO since $\alpha_s \mu_{\text{PS}} \sim
\delta$. Thus to convert the pole-scheme result for the renormalized
cross section to the PS scheme it is sufficient to replace $m_t^2-i
m_t \Gamma_t \rightarrow m_{t,\text{PS}}^2-i m_{t,\text{PS}} \Gamma_t
+\frac{\alpha_s}{\pi} \delta_1^{\text{PS}} \mu_{\text{PS}}
m_{t,\text{PS}}$ in the denominator of the top (antitop) propagator
(in both the LO and NLO cross section), $m_t \rightarrow m_{t,\text{PS}}$
in the numerator and to add the higher-order propagator contribution
\begin{equation}
\label{eq:prop-contr}
\delta {\cal A}^{(1)}_{\text{pole}\rightarrow\text{PS}} 
= {\cal A}^{(0)}  \frac{\frac{\alpha_s^2}{2 \pi^2} 
\delta_2^{\text{PS}} \mu_{\text{PS}} m_{t,\text{PS}}}
{p_t^2-m_{t,\text{PS}}^2+i m_{t,\text{PS}} \Gamma_t
-\frac{\alpha_s}{\pi} \delta_1^{\text{PS}} \mu_{\text{PS}} m_{t,\text{PS}}}
\end{equation} 
to the NLO amplitude and a similar one for the internal antitop propagator.

The result for the conversion from pole scheme to PS scheme can be
equivalently obtained considering how the top-quark propagator is
renormalized and resummed in the two schemes.  We introduce the
short-cut $D_t=p_t^2-m_t^2$ and
$D_{t,\text{PS}}=p_t^2-m_{t,\text{PS}}^2$ to indicate the renormalized
inverse propagator in the pole and PS scheme (without self-energy
resummation). In the pole scheme, the contribution of the hard part of
the one-loop QCD self-energy to the internal top-quark line can be
written as \cite{Falgari:2010sf}
\begin{equation} \label{eq:self_expansion}
\delta_{\text{QCD}}(m_t) 
\left[\frac{2 i m_t^2 (\dirac{p}_t+m_t)}{D_t^2}+
\frac{i m_t}{D_t}-\frac{i(\dirac{p}_t+m_t)}{D_t}\right] \, ,
\end{equation}   
with 
\begin{equation}
\delta_{\text{QCD}}(m_t)=\frac{\alpha_s C_F}{2 \pi} 
\left[\frac{3}{2 \epsilon}+2+\frac{x_{sc}}{2}
-\frac{3}{2} \ln\left(\frac{m_t^2}{\mu^2}\right)\right]
+{\cal O}(\alpha_s^2) \, ,
\end{equation}
while the correction to the top-quark line from the mass-renormalization
counterterm $\delta m_t$ reads
\begin{equation} \label{eq:self_expansion_b}
\frac{\delta m_t}{m_t} 
\left[\frac{2 i m_t^2 (\dirac{p}_t+m_t)}{D_t^2}+
\frac{i m_t}{D_t}\right] \, .
\end{equation}  
In the pole scheme one has $\delta m_t/m_t = -
\delta_{\text{QCD}}(m_t)$, which leads to the renormalized one-loop
propagator
\begin{equation}
\frac{i(\dirac{p}_t+m_t)}{D_t} (1-\delta_{\text{QCD}}(m_t) ) \, .
\end{equation}  
Including in addition the hard part of the electroweak self energy
one obtains
\begin{equation}
\frac{i(\dirac{p}_t+m_t)}{\Delta_t} (1-\delta_{\text{QCD}}(m_t) ) \, .
\end{equation}  

In the PS scheme the self-energy correction to the propagator has the
same functional form as in the pole scheme, but it depends on the
PS mass $m_{t,\text{PS}}$
\begin{equation} \label{eq:self_expansion_PS}
\delta_{\text{QCD}}(m_{t, \text{PS}}) 
\left[\frac{2 i m_{t,\text{PS}}^2 (\dirac{p}_t+ m_{t,\text{PS}})}{D_{t,\text{PS}}^2}+
\frac{i  m_{t,\text{PS}}}{D_{t,\text{PS}}}-\frac{i(\dirac{p}_t+
  m_{t,\text{PS}})}{D_{t,\text{PS}}}\right] \, . 
\end{equation}   
Analogously, the mass renormalization gives the contribution
\begin{equation} \label{eq:self_expansion_PS_b}
\frac{\delta m_{t,\text{PS}}}{m_{t,\text{PS}}} 
\left[\frac{2 i m_{t,\text{PS}}^2 (\dirac{p}_t+m_{t,\text{PS}})}{D_{t,\text{PS}}^2}+
\frac{i m_{t,\text{PS}}}{D_{t,\text{PS}}}\right] \, .
\end{equation}  
The mass counterterm in the PS scheme can be obtained using the
relation between the pole and PS mass, \Eqn{eq:PSmass}, and the
results given in Eqs.~(3.20) and (3.24) of Ref.~\cite{Falgari:2010sf}
$$
\frac{\delta m_{t, \text{PS}}}{m_{t, \text{PS}}}=
- \delta_{\text{QCD}}(m_{t,\text{PS}})+\frac{\mu_{\text{PS}}}{m_{t,
    \text{PS}}} \left[\frac{\alpha_s}{2 \pi}
  \delta_1^{\text{PS}}+\left(\frac{\alpha_s}{2 \pi} \right)^2
  \delta_2^{\text{PS}}+...\right] \, . 
$$
Thus the one-loop renormalized propagator in the PS scheme reads
\begin{eqnarray} \label{eq:self_expansion_PS_c}
&&\frac{i(\dirac{p}_t+ m_{t,\text{PS}})}{D_{t,\text{PS}}}
  (1-\delta_{\text{QCD}}(m_{t,\text{PS}}))\nonumber\\ &&
 +\frac{\mu_{\text{PS}}}{m_{t,\text{PS}}}
  \left[\frac{\alpha_s}{2 \pi}
    \delta_1^{\text{PS}}+\left(\frac{\alpha_s}{2 \pi}\right)^2
    \delta_2^{\text{PS}}\right] \left(\frac{2 i m_{t,\text{PS}}^2
    (\dirac{p}_t+ m_{t,\text{PS}})}{D_{t,\text{PS}}^2}
   + \frac{i m_{t,\text{PS}}}{D_{t,\text{PS}}}\right) \, .
\end{eqnarray} 
Remembering that $\mu_{\text{PS}} \sim \alpha_s m_t$ and keeping only
the parametrically relevant terms one obtains
\begin{equation} \label{eq:self_expansion_PS_d}
\frac{i(\dirac{p}_t+ m_{t,\text{PS}})}{D_{t,\text{PS}}}
\left(1-\delta_{\text{QCD}}(m_{t,\text{PS}}) 
+ \frac{\frac{\alpha_s}{\pi} \delta_1^{\text{PS}}  \mu_{\text{PS}}
  m_{t,\text{PS}}  +\frac{\alpha_s^2}{2 \pi^2} \delta_2^{\text{PS}}
  \mu_{\text{PS}} m_{t,\text{PS}}}{D_{t,\text{PS}}}  \right)\, . 
\end{equation} 
The terms proportional to $\delta_{\text{QCD}}$ and
$\delta_2^\text{PS}$ scale as $\sim \alpha_s \sim \sqrt{\delta}$
compared to the leading propagator, and can be included perturbatively in the
calculation of the cross section. In contrast, the
term proportional to $\delta_1^\text{PS}$ is a correction of order
one, since $\alpha_s \mu_\text{PS} m_{t,\text{PS}}/D_{t,\text{PS}} \sim 1$.
It thus has to be resummed into the leading-order
propagator. Including the contribution of the one-loop
electroweak self energy, the resummed renormalized propagator reads
\begin{eqnarray} \label{eq:self_expansion_PS_e}
&&\frac{i(\dirac{p}_t+
    m_{t,\text{PS}})}{\Delta_{t,\text{PS}}-\frac{\alpha_s}{\pi}
    \delta_1^{\text{PS}}  
\mu_{\text{PS}} m_{t,\text{PS}} } \left[1-\delta_{\text{QCD}}(m_{t,\text{PS}})
+  \frac{\frac{\alpha_s^2}{2 \pi^2}    \delta_2^{\text{PS}}
  \mu_{\text{PS}} m_{t,\text{PS}}
}{\Delta_{t,\text{PS}}-\frac{\alpha_s}{\pi} \delta_1^{\text{PS}}  
\mu_{\text{PS}} m_{t,\text{PS}} }  \right] \, ,
\end{eqnarray} 
with $\Delta_{t,\text{PS}}=p_t^2-m_{t,\text{PS}}^2+i m_{t,\text{PS}} \Gamma_t$.
This confirms that the conversion from pole to PS scheme amounts to 
the replacement $\Delta_t \rightarrow \Delta_{t,\text{PS}}
-\frac{\alpha_s}{\pi} \delta_1^{\text{PS}} \mu_{\text{PS}}
m_{t,\text{PS}}$ in the denominator of the resummed top-quark
propagator and the addition of the term
\begin{equation}
\delta {\cal A}^{(1)}_{\text{pole}\rightarrow\text{PS}} = 
{\cal A}^{(0)}   \frac{\frac{\alpha_s^2}{2 \pi^2} 
\delta_2^{\text{PS}} \mu_{\text{PS}} m_{t,\text{PS}}   }{\Delta_{t,\text{PS}}
-\frac{\alpha_s}{\pi} \delta_1^{\text{PS}} 
\mu_{\text{PS}} m_{t,\text{PS}} }
\end{equation} 
to the NLO amplitude for each intermediate top or antitop quark line
present at LO.

\subsection{Validity of EFT results}
\label{sec:valEFT}

We conclude this section with a remark on the validity of the EFT
results for $t \bar{t}$ production. As outlined in
Section~\ref{sec:EFT}, in our effective-theory framework the unstable
particle is correctly treated by HQET (plus finite-width
effects). Strictly speaking, for production of a pair of unstable
particles this is true as long as the two velocities $v$, $\bar{v}$
are relativistic, which is the case if $\sqrt{\hat{s}} \gg 2 m_t$,
namely when the partonic centre-of-mass energy $\sqrt{\hat{s}}$ is much
larger than the pair-production threshold. When $\sqrt{\hat{s}} - 2
m_t \sim \Gamma_t$ the two particles become non-relativistic, $v \sim
\bar{v} \sim (1,\vec{0})$, and the correct effective theory to
describe the pair-production process is non-relativistic QCD
(NRQCD). In the expansion by regions this is seen as an additional
\emph{potential} region which develops when the heavy particles become
non-relativistic, and describes Coulomb interactions of the two
slow-moving particles.  This breakdown of the expansion in $\delta$
was already noticed in the context of the pole approximation for
$W$-pair production, with the \emph{double-pole approximation} quickly
loosing accuracy when $\sqrt{\hat{s}}$ approaches $2 M_W$
\cite{Denner:2000bj}. While this is relevant for an $e^- e^+$
collider, where the centre-of-mass energy can be tuned to probe the
threshold region, at hadron colliders the physical hadronic cross
section is obtained by convoluting the partonic cross section and
parton luminosities over the full range
$\sqrt{\hat{s}}=[0,\sqrt{s}]$. For a multi-TeV hadronic c.o.m. energy
$\sqrt{s}$, and particle masses in the range of 100-200 GeV, the
contribution of the threshold region is expected to be small, so that
the error introduced by using the wrong EFT description near threshold
is negligible. We have checked this explicitly for the total NLO cross
section in the stable-top approximation and found that the threshold
region, defined by $v \le$ 0.1-0.2, contributes 1-5\% of the total
hadronic cross section. The Coulomb terms, which are the ones 
incorrectly reproduced by HQET in the threshold limit, are in fact less
than 1\%.

Below threshold, doubly-resonant configurations do not dominate the
cross section because the phase space for production of two resonant
tops is strongly suppressed for $\sqrt{\hat{s}}<2 m_t$. In this region
NRQCD should be replaced by a new effective theory describing
singly-resonant production of a $t \bar{t}^*$ ($t^* \bar{t}$) pair,
which would require the calculation of non-trivial QCD corrections to
the processes $q \bar{q} \rightarrow t W^- \bar{b} (\bar{t} W^+ b)$.
For this reason, below threshold we switch to a Born-level prediction
of the top-pair production cross section, and set virtual and real
corrections to zero. Once again, one expects the error in introducing this
approximation to be below the accuracy of a few percent which we pursue here.
Actually, in the results presented next, we impose a physical cut on 
the final state so that we never run into the complications near threshold. 

\section{Results}
\label{sec:res}

In this section we present explicit results for a sample of
representative kinematical distributions relevant for top quark pair
production studies. Our main purpose here is to discuss the importance of
off-shell and factorizable corrections by making a direct comparison to 
the NWA approach. The
distributions presented in Sections~\ref{sec:inv_and_tr_masses} -
\ref{sec:fb_asymmetry} are computed in the pole-mass renormalization
scheme whilst a comparison to results obtained in the PS scheme is
given in Section~\ref{sec:PSvsPole}.

The input values for the SM parameters are given in Table
\ref{table:tevatron-setup}. For the convolution of the partonic cross
section with the PDFs we use the MSTW2008 NLO set~\cite{Martin:2009iq} for both
LO and NLO cross sections, taking the corresponding value for $\alpha_s$ 
for consistency. In addition, for
all contributions involving the top decay width we use the NLO width,
$\Gamma^{\text{NLO}}_t$. Using NLO PDFs and $\Gamma^{\text{NLO}}_t$
throughout is a choice that has been made to ease identification of
off-shell effects, ensuring that corrections arising from changes in
PDFs or decay widths do not cloud the effects introduced by the
off-shellness of the tops.  Our default choice of factorization and
renormalization scales is $\mu_F=\mu_R=m_t$. Where present, bands
around LO and NLO off-shell results are obtained by simultaneously
varying the renormalization and factorization scales in the interval
$[m_t/2,2 m_t]$.

\begin{table}
\centering
\begin{tabular}{c c c}
\hline \\[-6pt]
\multicolumn{3}{c}{
Collider: Tevatron, $\sqrt{s} = 1.96 \; \text{TeV}$ } \\[5pt]
\hline \hline \\[-6pt]
$p_T(J_b) > 15 \; \text{GeV}$ & $p_T(l^+) > 15 \; \text{GeV}$ & $\slashed{E}_T > 20 \; \text{GeV}$ \\[5pt]
$p_T(J_{\bar{b}}) > 15 \; \text{GeV}$ & $p_T(l^-) > 15 \; \text{GeV}$ & $D_{\text{jet}} = 0.7$\\[5pt] 
\hline \\[-6pt]
$M_t = 172.9 \; \text{GeV}$ & $\Gamma^{\text{NLO}}_t = 1.3662 \; \text{GeV}$ & $M_Z = 91.2 \; \text{GeV}$ \\[5pt] 
$M_W = 80.4 \; \text{GeV}$ & $\Gamma_W = 2.14 \; \text{GeV}$ & $\alpha_{\text{ew}} = 0.03394$ \\[5pt] 
\hline \hline
\end{tabular}
\caption{Example process definition at the Tevatron collider and
  parameter setup.}
\label{table:tevatron-setup}
\end{table}
 
Throughout the following discussion we assume that the final state
contains a $b$ and a $\bar{b}$ jet and furthermore that the $W$-bosons
are perfectly reconstructed. Despite the latter not being feasible in
the di-lepton channel it allows one to identify features arising from
the off-shellness of the top quarks that will also be relevant in an
experimentally rigorous analysis. It is emphasised however, that
wherever we discuss an observable related to the (anti)top, it is
understood this is the \emph{reconstructed} (anti)top quark, defined in
terms of the $W^{+(-)}$-boson and ($\bar{b}$)$b$-jet,
$J_{b(\bar{b})}$.

In order that the EFT approach and counting remains valid the following
conditions on the reconstructed top-quark momenta, $p(t) =
p(J_b)+p(W^+)$ and $p(\bar{t}\,)=p(J_{\bar{b}})+p(W^-)$, are also
imposed:
\begin{align}
140 \;\; &\text{GeV} \;<\: M_{\text{inv}}(t) = \sqrt{\left( p(t)
  \right)^2} \:<\: 200 \;\; \text{GeV}, \nonumber \\[10pt] 
140 \;\; &\text{GeV} \;<\: M_{\text{inv}}(\bar{t}\,) = \sqrt{\left(
  p(\bar{t}) \right)^2} \:<\: 200 \;\; \text{GeV}, \nonumber \\[10pt] 
& \hspace{-0.1cm} M_{\text{inv}}(t\bar{t}\,) = \sqrt{\left(p(t) +
  p(\bar{t} \,) \right)^2} \:>\: 350 \;\; \text{GeV}. 
\end{align}   
The experimental setup we consider, typical for a study of top-pair
production, is detailed in Table \ref{table:tevatron-setup} and
defines the process
\begin{align}
p \: \bar{p} \: \rightarrow \: J_b \: J_{\bar{b}} \: \slashed{E}_T \: l^+ \: l^- + X.
\end{align}
We note that the jets have been clustered using the standard
$k_t$-algorithm with jet resolution parameter $D_{jet}=0.7$.

The implementation of the factorizable corrections to production and
decay subprocesses has been checked in the limit of on-shell top
quarks against the di-lepton channel implementation in MCFM
\cite{Campbell:2012uf}. Very good agreement was found both for inclusive 
cross sections and for all distributions we have checked. The analytical expressions for the
virtual non-factorizable corrections have been put through extensive
numerical checks and corresponding real corrections display the
correct independence on the soft FKS parameter $\xi_{\text{cut}}$.

\subsection{Invariant and Transverse Masses}
\label{sec:inv_and_tr_masses}

\begin{figure}[t!]
\centering
\includegraphics[trim=1.2cm 0.0cm 3.0cm 0.3cm,clip,width=12.0cm]{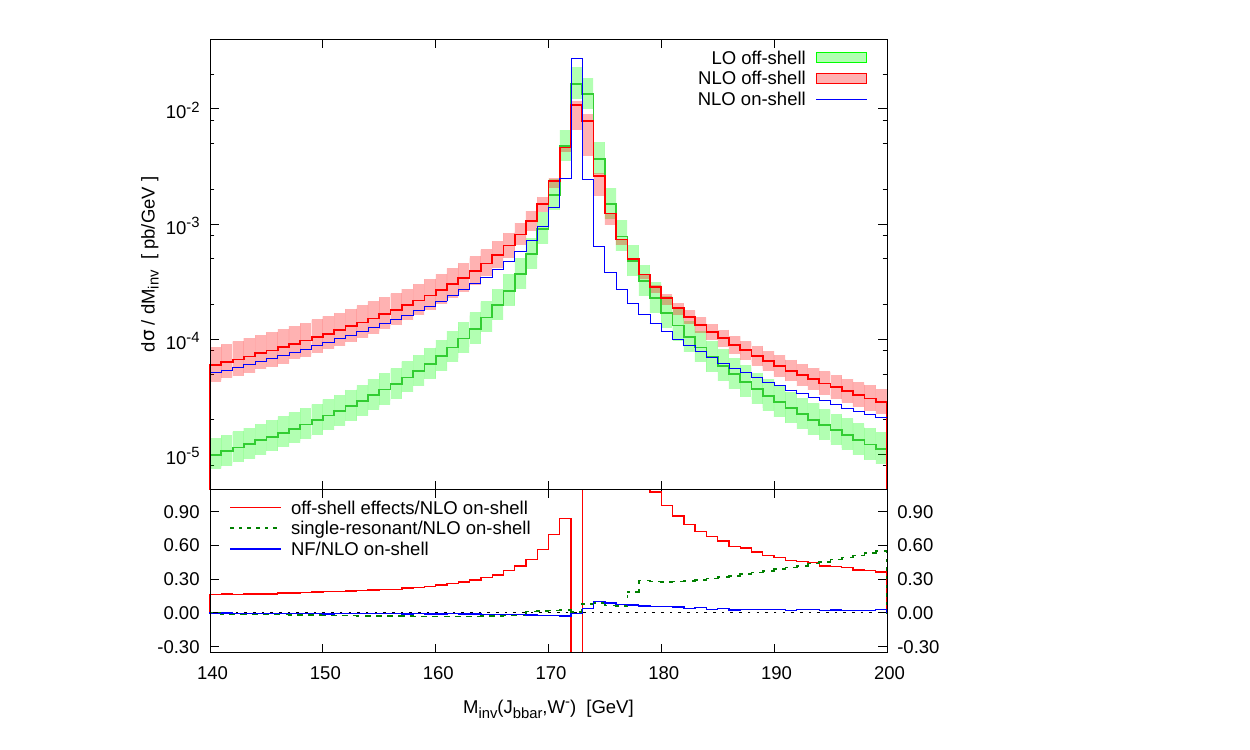}
\caption{Invariant mass distribution for the reconstructed antitop
  quark.  Upper plot: results for the off-shell LO result (green),
  off-shell NLO result (red) and on-shell NLO prediction (blue). Lower
  inset: relative contributions of total off-shell effects,
  non-factorizable corrections and singly-resonant diagrams. See the
  text for explanation.}
\label{fig:inv_mass}
\end{figure}

\begin{figure}[t!]
\centering
\includegraphics[trim=1.0cm 0.0cm 3.0cm 0.3cm,clip,width=12.0cm]{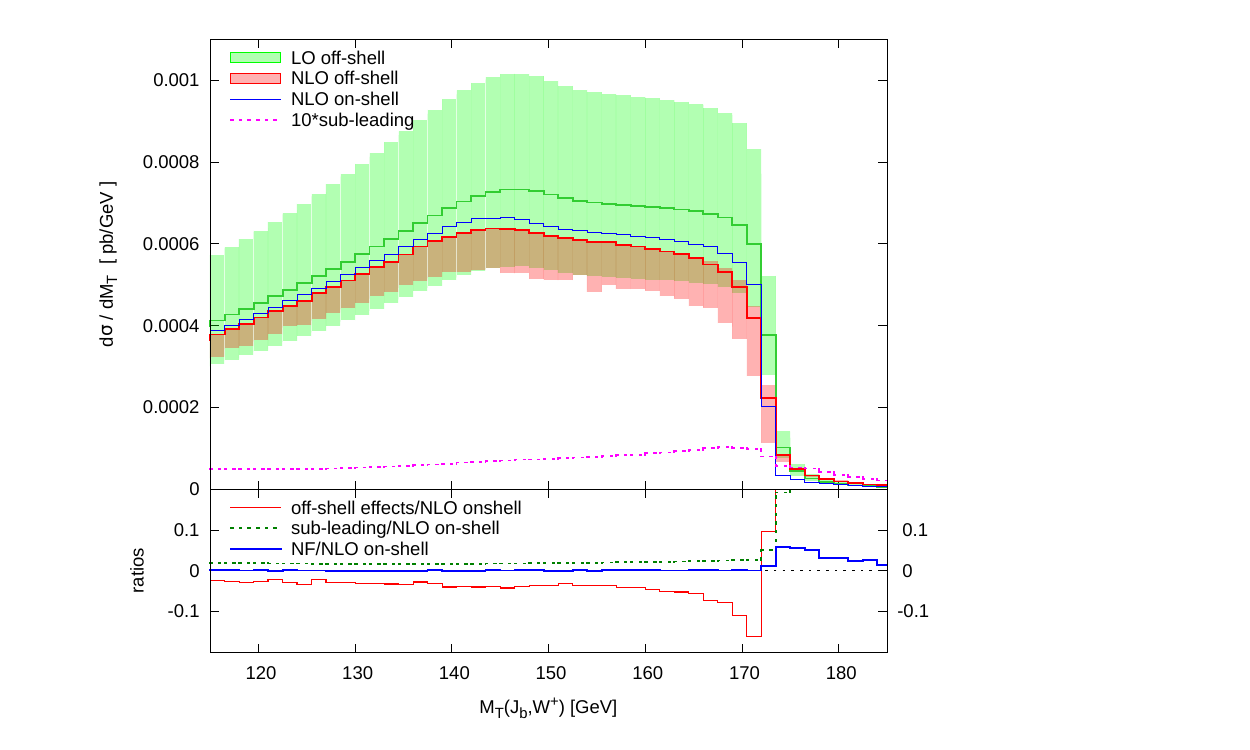}
\caption{Transverse mass distribution for the reconstructed top quark.
  Upper plot: results for the off-shell LO result (green), off-shell
  NLO result (red) and on-shell NLO prediction (blue).  Lower inset:
  relative contributions of total off-shell effects, non-factorizable
  corrections and singly-resonant diagrams.  See the text for
  explanation.}
\label{fig:transverse_mass}
\end{figure}

We start by presenting results for the invariant and transverse masses
of the reconstructed tops, displayed in Figures \ref{fig:inv_mass} and
\ref{fig:transverse_mass} respectively. These observables are expected
to be sensitive to the off-shellness of the top quarks. The off-shell
LO (green band) and NLO (red band) results are displayed in the upper
plots, as is the NLO prediction in the NWA (blue, solid). In order to
straightforwardly assess the importance of off-shell effects,
non-factorizable corrections and sub-leading contributions, we study
the quantities
\begin{align}
\frac{\sigma^{\text{NLO, off-s.}}-\sigma^{\text{NLO,
      on-s.}}}{\sigma^{\text{NLO, on-s.}}} &\;\;\; \text{(red,
  solid)}, \label{eq:off-shell-effects_ratio} \\[10pt]  
\frac{\sigma^{\text{NF}}}{\sigma^{\text{NLO, on-s.}}} \hspace{1.0cm}
&\;\;\; \text{(blue, solid)}, \label{eq:nf_ratio} \\[10pt] 
\frac{\sigma^{\text{sub-lead.}}}{\sigma^{\text{NLO,on-s.}}} 
\hspace{1.0cm} &\;\;\; 
\text{(green, dashed)}, \label{eq:sub-leading_ratio}  
\end{align}   
in the lower panel of each figure. The non-factorizable corrections
have been evaluated at the scale $\mu_{\text{soft}} \simeq m_t \,
\delta$, as motivated at the end of Section~\ref{sec:virtual}. For the
case at hand this amounts to multiplying $\sigma^{\text{NF}}(\mu=m_t)$
by the factor $\alpha_s(\mu_{\text{soft}})/\alpha_s(m_t) \simeq 2.24$,
when $\delta = 0.02$.  It would also be desirable to include a
resummation of logarithms of $\mu_{\text{soft}}/m_t$ by running the
hard matching coefficients down to the soft scale using RGEs, however
this will not be discussed here.

Firstly we focus on the invariant mass distribution of the reconstructed
antitop quark, shown in Figure \ref{fig:inv_mass}.\footnote{The LO
  on-shell result is a delta function centred at $m_t$ and it is
  omitted in the plot.}  An important feature present in the upper
plot is that the shapes of the NLO off-shell and on-shell
distributions are significantly different, particularly near and
beyond the peak position.  This is made clear in the lower inset where
one observes that off-shell effects are indeed sizeable, exceeding $60\%$
around the peak.  However, one also notices that they change sign in
this region, resulting in the net off-shell effects being small due to an
averaging out that occurs for inclusive enough observables like the
total cross section.  The lower inset in Figure \ref{fig:inv_mass}
also shows the effect of strict non-factorizable effects and
sub-leading contributions, both normalized to the NLO on-shell results.  
Non-factorizable corrections, given by the sum of virtual
and real soft contributions, are very small, $\sim$ 1\% over
the whole invariant-mass range considered, except around the pole, where
they grow to roughly 6-7\%. The contribution of
sub-leading diagrams is also very small close to the peak, where
$\delta \ll 1$ and the effective-theory counting is satisfied, but 
becomes sizeable in the tail region for $M_{\text{inv}}(t) \,> \,m_t$, as one would 
expect.
 
We now consider Figure \ref{fig:transverse_mass} which displays the
transverse mass distribution of the reconstructed top quark, defined
via
\begin{align}
M_T(t) &= \left((|\vec{p}_T(J_b)| + |\vec{p}_T(l^+)| + | E_T(\nu_l)|)^2 -(\vec{p}_T(J_b) + \vec{p}_T(l^+) + \vec{p}_T(\nu_l))^2 \right)^{1/2}  
\end{align}
where $\vec{p}_T(k)$ is the transverse momentum of the final state $k$.

Once more, it is clear that there are substantial differences between
the shape of the distribution in the on-shell and off-shell
approaches.  In particular, the relatively sharp edge in the on-shell
calculation becomes much less steep once the on-shell assumption
is relaxed.  The size of off-shell effects are quantified in the lower
inset where one observes that these constitute a negative 2-3\%
correction to the NWA result, except near and beyond the edge (at $M_T
\sim m_t$) where much larger effects are present. In this region, the
averaging effect mentioned earlier is less effective leading to
enhanced corrections.  The pure non-factorizable corrections are
negligible for almost the entire range, but do grow to 5-6\% near the
edge.  Finally it is clear from the figure that the sub-leading terms
are also small ($\sim 2\%$) for values of $M_T$ below $m_t$ and become
more important for values beyond the edge.

The common characteristics shared by the contributions
(\ref{eq:off-shell-effects_ratio}) - (\ref{eq:sub-leading_ratio}) for
this observable can be understood by the fact that the distribution in
the on-shell approach has a fixed edge at $m_t$ at LO, whilst at NLO
it can only receive contributions beyond $m_t$ from events involving
additional radiation that is clustered into $J_b$. In the off-shell
computation (at both LO and NLO) this region receives additional
contributions from events where $M_{\text{inv}}(t) > m_t$ and thus
differs from the on-shell result distinctly. It should be noted that
in this region soft-gluon resummation or parton shower effects are
also expected to play an important role.

\subsection{Individual NLO contributions}
\label{sec:breakdown_of_contributions}

\begin{figure}[t!]
\centering
\includegraphics[trim=1.0cm 0.0cm 3.0cm
  0.7cm,clip,width=12.0cm]{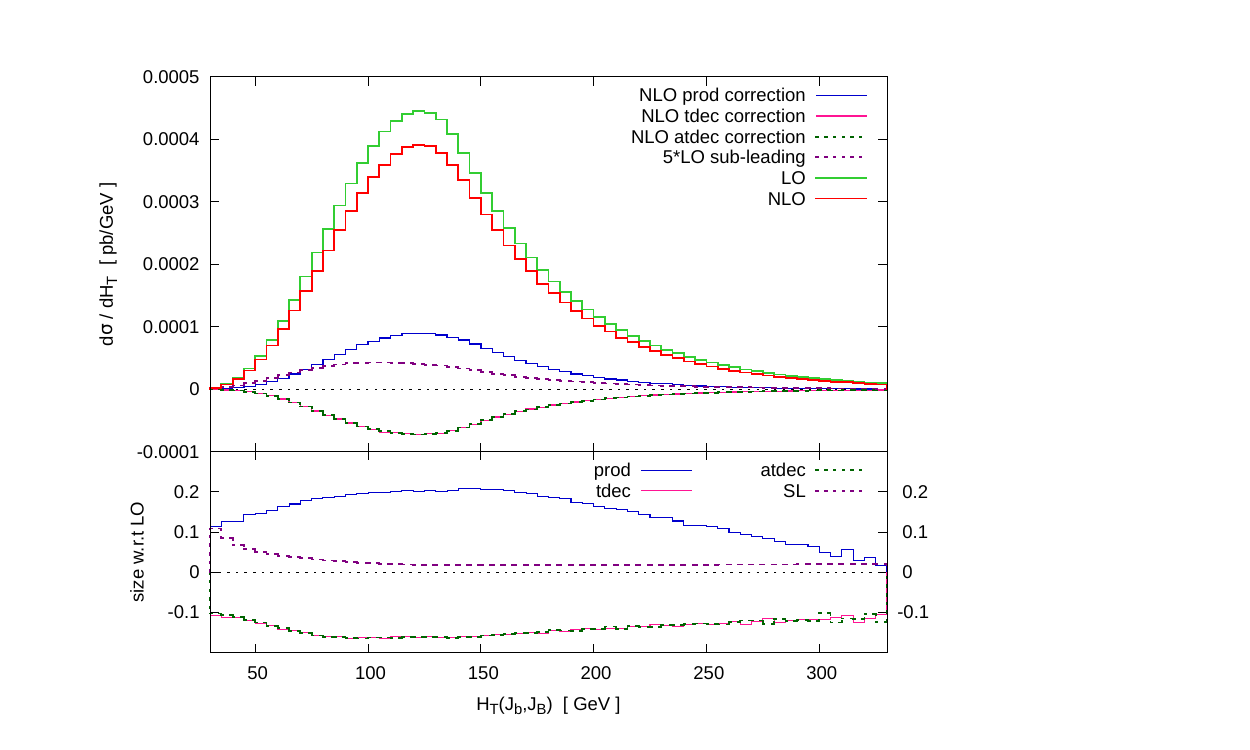} 
\caption{$H_T(J_b, \, J_{\bar{b}})$ distribution.  Upper plot:
  off-shell LO (green, solid) and NLO (red, solid) results. The
  factorizable corrections to the production (blue, solid), top decay
  (pink, solid) and antitop decay (dark green, dashed) subprocesses
  are also displayed along with the tree-level sub-leading
  contributions (purple, dashed). The ratios of the latter four
  contributions with respect to the LO results are shown in the lower
  inset (same colours and line-styles).  See the text for
  explanation.}
\label{fig:separate_contributions_HT_JbJB}
\end{figure}

We now examine the relative importance of the different contributions
to the cross section for the distribution $H_T(J_b, \, J_{\bar{b}}) =
p_T(J_b)+p_T(J_{\bar{b}})$. The findings, illustrated in Figure
\ref{fig:separate_contributions_HT_JbJB}, are typical of observables
that are inclusive in the invariant masses of the top quarks.

It is clear from the figure that the corrections to the top and
antitop decay subprocesses are important for the correct
normalization of the distribution, each separately correcting the LO
results by about $-10$\%. The negative sign of the latter means that
there is a partial cancellation of NLO corrections between the
production and decay subprocesses. It is evident from the almost flat
shape of the decay corrections in the lower inset of Figure
\ref{fig:separate_contributions_HT_JbJB} that these do not noticeably
alter the shape of the distribution.  Rather, any significant
correction to the LO shape comes about through the factorizable
corrections to the production subprocess. The non-factorizable
corrections have not been included in the plot as they are tiny, as
expected by the almost complete cancellation between real and virtual
corrections for inclusive observables.  In fact, the off-shell effects
(also not displayed) are, as a whole, also small over the full range of
$H_T$ considered, in agreement with the a priori expectation that
these effects are of order $\sim \Gamma_t/m_t$.  The sub-leading
contributions constitute a 1-2\% correction to the LO result (except
at very low $H_T$), indicating that the EFT power-counting is working
well.

\subsection{Forward-Backward Asymmetry}
\label{sec:fb_asymmetry}

 \begin{figure}[t!]
 \centering
 \includegraphics[trim=1.0cm 0.0cm 3.0cm 1.05cm,clip,width=12.0cm]{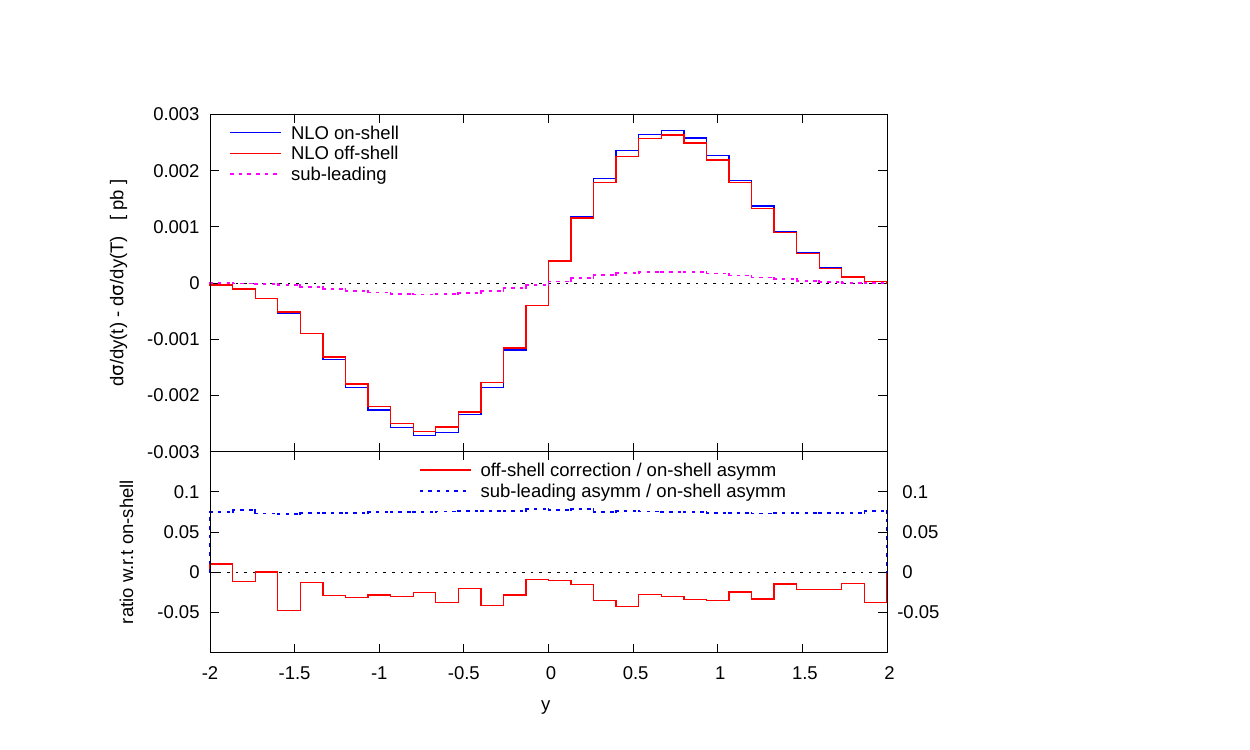}
 \caption{Differential forward-backward asymmetry for the
   reconstructed top quarks, see Eq. (\ref{eq:differential-fb-asymmetry}).  
   Upper plot: off-shell (red, solid),
   on-shell (blue, solid) and sub-leading (purple, dashed)
   asymmetries.  Lower inset: ratio of the off-shell correction (red,
   solid) and sub-leading asymmetries (blue, dashed) with respect to
   the on-shell asymmetry. See text for further explanation. }
 \label{fig:fb_asymm_top}
 \end{figure}

Figure \ref{fig:fb_asymm_top} displays the differential distribution
relevant for the study of the forward-backward asymmetry of the top
quark,
\begin{align}
\text{top asymmetry: } \; &\frac{d\sigma}{dy(t)} -
\frac{d\sigma}{dy(\bar{t}\,)} .  
\label{eq:differential-fb-asymmetry}
\end{align}
The important feature to pick out is that off-shell effects amount to
a small negative correction to the on-shell asymmetry of less than 4\%
in magnitude. This is fully expected for an observable that is
inclusive over the invariant mass of the top.  The sub-leading terms
display a small asymmetry, giving a $\sim 7\%$ correction to the
on-shell asymmetry.

\subsection{Pole mass versus PS mass}
\label{sec:PSvsPole}

In this subsection we examine the effects of using the PS-mass scheme
as an alternative to the pole scheme. As discussed previously, the PS
mass has the advantage of being free of non-perturbative ambiguities
whilst still being a suitable mass for the effective-theory setup as
long as the choice $\mu_{\text{PS}} \sim \alpha_s m_t$ is made.

\begin{figure}[h!]
 \centering
 \includegraphics[trim=0.3cm 0.3cm 0.2cm
   0.4cm,clip,width=16.5cm]{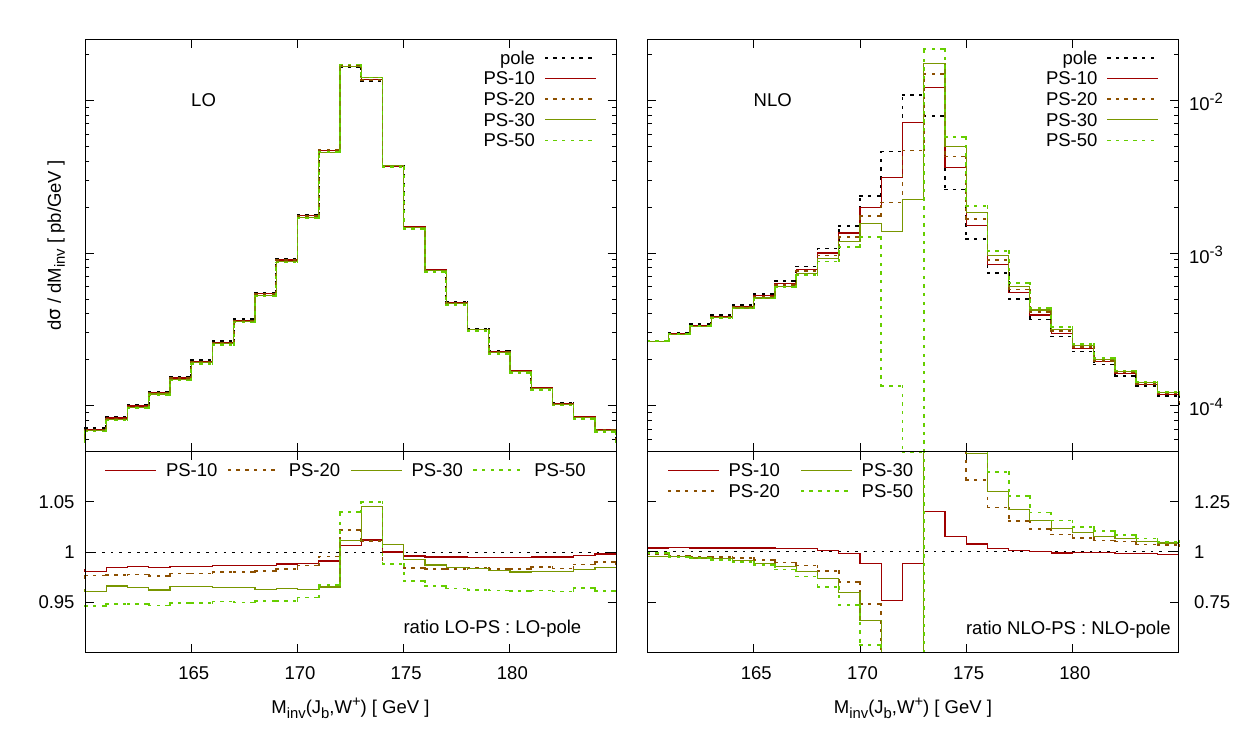}  
 \caption{Invariant-mass distribution for reconstructed top quark at LO
   (left panel) and NLO (right panel) for pole and various
   PS-schemes. See the text for explanation.}
 \label{fig:ps_minv1}
 \end{figure}

\begin{figure}[h!]
 \centering
 \includegraphics[trim=0.3cm 0.3cm 0.2cm
   0.4cm,clip,width=16.5cm]{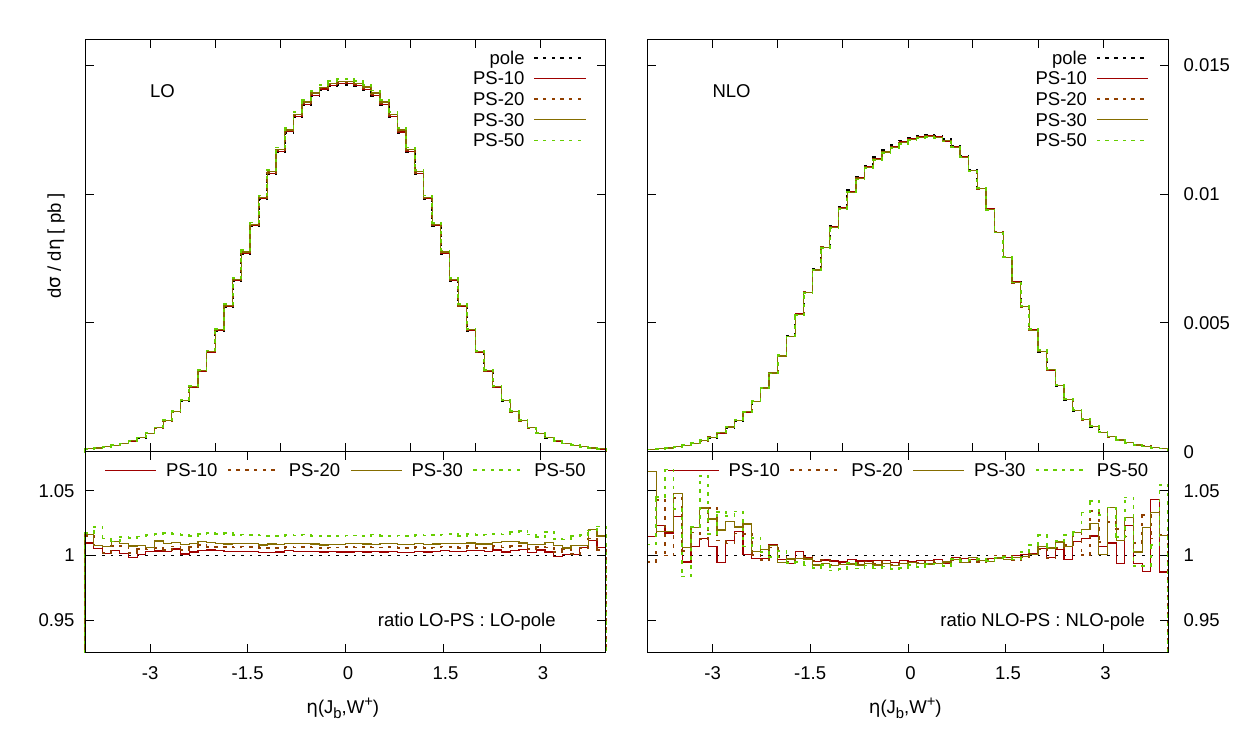}  
 \caption{Pseudo-rapidity distribution for reconstructed top quark at LO
   (left panel) and NLO (right panel) for pole and various
   PS-schemes. See the text for explanation.}
 \label{fig:ps_eta}
 \end{figure}

We illustrate the above statements in Figures \ref{fig:ps_minv1} and
\ref{fig:ps_eta} where the reconstructed top-quark invariant mass
and pseudo-rapidity are shown respectively. Distributions in the pole
scheme and PS-scheme with $\mu_{\text{PS}}\in \{10,20,30,50 \}$~GeV at
LO (LHS panels) and NLO (RHS panels) are plotted. In detail, we fix
the numerical value of $m_{t,\text{PS}}(\mu_{\text{PS}})$ to the pole
mass $m_t=172.9~\text{GeV}$ at next-to-leading order, i.e. including
the term $\sim \delta_1^{\text{PS}}$ in \Eqn{eq:PSmass} and obtain
$m_{t,\text{PS}}(10)=172.44~\text{ GeV}$,
$m_{t,\text{PS}}(20)=171.97~\text{GeV}$,
$m_{t,\text{PS}}(30)=171.53~\text{GeV}$ and
$m_{t,\text{PS}}(50)=170.6~\text{GeV}$ respectively. The width of the
top is also adapted in accordance with $m_t$ at next-to-leading order
in $\alpha_s$.

According to our discussion in Sections~\ref{sec:EFT_mass-scheme} and
\ref{sec:mass}, we would expect the suitability of
$\mu_{\text{PS}}=50$~GeV to be highly questionable, whereas
$\mu_{\text{PS}}=10$~GeV should be perfectly acceptable. To confirm
this, firstly, we point out that at LO the distributions in all
schemes agree very well with each other (within $\sim$ 5\% for all
bins), even for `bad' scheme choices. This is clear since terms
affecting the perturbative stability of a particular mass scheme only
enter at orders beyond LO. In the RH panel of Figure \ref{fig:ps_minv1}, 
where the NLO invariant mass distributions are plotted, it is evident that the
choice $\mu_{\text{PS}} \gtrsim 30$~GeV leads to serious deviations in
shape from the pole-scheme curve and to large NLO corrections. This
shift is due to the term given in \Eqn{eq:prop-contr} and, to a
lesser extent, due to non-factorizable corrections.  The size of the
NLO corrections implies that a NLO description of the lineshape for
$\mu_{\text{PS}} \gtrsim 30$~GeV is not trustworthy. In the same plot
however, we observe that sensible scheme choices ($\mu_{\text{PS}} =
10, \, 20$ GeV) give lineshapes that have a stable perturbative
expansion. Thus it is perfectly legitimate to use such schemes rather
than the pole-mass scheme for the extraction of the top mass from a
measurement of the invariant mass of its decay products.
Moreover, at NLO it is only observables
that are more exclusive in the invariant mass that will display
significant differences in shape for different schemes. The latter
feature is illustrated well for the top pseudo-rapidity in the RH
panel of Figure \ref{fig:ps_eta}, where all schemes display the same
shape and the NLO corrections are under control.

 \begin{figure}[h!]
 \centering
 \includegraphics[trim=1.0cm 0.0cm 3.0cm
   0.4cm,clip,width=12.0cm]{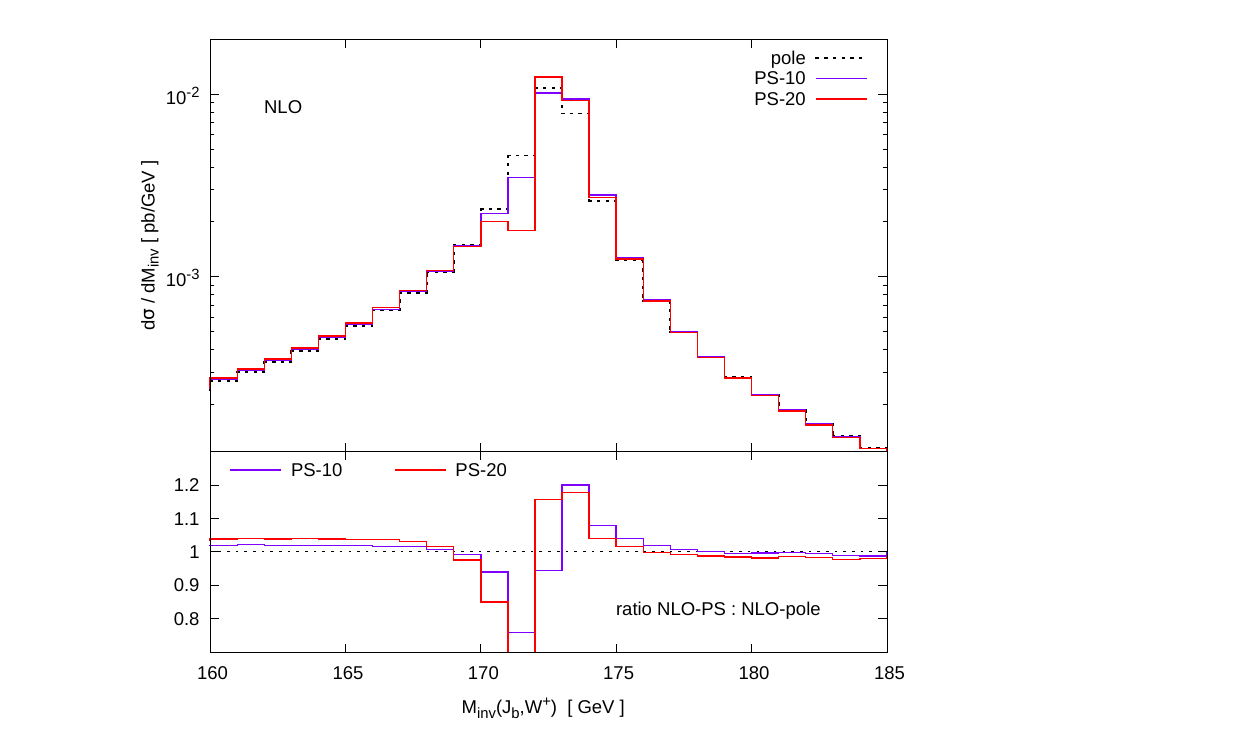}  
 \caption{Invariant-mass distribution for reconstructed top quark at NLO for
   pole and various PS-schemes. See the text for explanation. } 
 \label{fig:ps_minv-correct}
 \end{figure}

In order to investigate the impact of mass-scheme ambiguities on the
determination of the top mass, we now consider in a toy analysis 
the invariant mass distribution for
$\mu_{\text{PS}}\in\{0,10,20\}$~GeV, where $\mu_{\text{PS}}=0$ of
course corresponds to the pole scheme. We want to compare the
extraction of the top mass in these schemes at LO and NLO. We stress
that we only consider effects from scheme changes and ignore all other
effects, such as colour-reconnection effects and many more.

Starting at LO, we extract the mass in all schemes by adjusting its
value to obtain optimal agreement with the measured distribution. Let us
assume in the pole scheme the measured distribution is matched
perfectly for $m_t = 172.9$~GeV. After the extraction of this value
for the pole mass we can now convert the latter to the $\overline{\rm
  MS}$ scheme to obtain $\overline{m}=162.2$, where the conversion is
done at three loop~\cite{Melnikov:2000qh} with a crude Pade
estimate\footnote{We emphasise that the precise method employed to
  estimate effects of higher order corrections does not play a major
  role here. For all numbers quoted in this subsection, including
  Table \ref{table:mass-extraction}, the estimate of the error in the
  conversion is $\lesssim 100$~MeV. } for higher-order
corrections. However, as we have argued above, we are also entitled to
use the PS scheme with $\mu_{\text{PS}}=10$~GeV or
$\mu_{\text{PS}}=20$~GeV for such an analysis. Given the results shown
in the left panel of Figure~\ref{fig:ps_minv1}, the extracted values
for the masses in these schemes are extremely close to the NLO
converted masses given previously in this subsection and listed in the first column of
Table~\ref{table:mass-extraction}. If we now convert these values to
the $\overline{\rm MS}$ scheme with precisely the same procedure used
for the pole scheme, we obtain the values for $\overline{m}$ listed in
the second column of Table~\ref{table:mass-extraction}. These values
differ by $600-800$~MeV from $\overline{m}=162.2$. Alternatively, if
we convert the extracted values of
$m_{t,\text{PS}}(10)=172.44~\text{GeV}$ and
$m_{t,\text{PS}}(20)=171.97~\text{GeV}$ back to the pole scheme, we
obtain the values $m_t=173.5$~GeV and $m_t=173.8$~GeV respectively, as
listed in the third column of Table~\ref{table:mass-extraction}. This
conversion is done at ${\cal O}(\alpha_s^4)$, again supplemented by a
crude Pade approximation to estimate higher-order effects. We stress
that all values given in the left panel of
Table~\ref{table:mass-extraction} are of course consistent at ${\cal
  O}(\alpha_s)$ and differ only by terms that are formally of higher
order. The crucial point is that the numerical effect of these
higher-order terms lead to an ambiguity of $500-900$~MeV in the LO
extraction of the top mass from the invariant mass distribution.

\begin{table}
\centering
\begin{tabular}{c||ccc||ccc}
\hline 
& & & & & &  \\[-9pt]
 & \multicolumn{3}{c||}{ LO}  & \multicolumn{3}{c}{NLO} \\
$\mu_{\text{PS}}$ & $m_{\text exp}$ & $\overline{m}$ & $m_t$ \qquad
& $m_{\text exp}$ & $\overline{m}$ & $m_t$ \\[3pt]
\hline \hline 
& & & & & &  \\[-9pt]
0 & 172.9 & 162.2 & 172.9 & 172.9 & 162.2 & 172.9 \\
10 & 172.4 & 162.7 & 173.5 & 172.2 & 162.4 & 173.3 \\
20 & 172.0 & 163.0 & 173.8 & 171.5 & 162.5 & 173.4 \\[3pt]
\hline 
\end{tabular}
\caption{Extraction of the top mass in various schemes at LO (left
  panel) and NLO (right panel). All numbers are in units of GeV. }
\label{table:mass-extraction}
\end{table}

We can now repeat this exercise at NLO where we begin by stressing
that NLO in this context implies the inclusion of NLO corrections to
the propagation of the top quark, in addition to the factorizable and
non-factorizable corrections. Simply taking into account NLO
corrections to the production subprocess alone does not improve the LO
toy analysis presented above.  We again assume that the experimental
measurement of the invariant mass is perfectly matched in the pole
scheme by setting $m_t=172.9$~GeV. To extract the PS-mass from this
measurement we would have to perform a best-fit analysis in the
PS-scheme and extract the best value for the top mass. We have found
that this best value is very close (but not exactly identical) to the
value obtained from two-loop conversion of the pole mass to the PS
mass, listed in the first column in the right panel of
Table~\ref{table:mass-extraction}. In Figure~\ref{fig:ps_minv-correct}
we show the corresponding distributions to confirm that they are
indeed very close.
 
As in the LO case we then convert the extracted values for the mass to
the $\overline{\rm MS}$ scheme (second column) and back to the pole
scheme (third column). Not surprisingly, the ambiguities in the
extraction of the mass has decreased compared to the LO case and is now
about $200-300$~MeV for the determination of $\overline{m}$ and
$400-500$~MeV for the determination of the pole mass. 

Let us conclude this subsection with a few comments regarding this toy
analysis. First and foremost, this is of course only a basic 
investigation. Apart from taking into account all partonic channels, a
fully rigorous analysis would have to include numerous effects beyond fixed
order in perturbation theory. However, the main result that there is
an existing additional ambiguity due to the scheme dependence, is unlikely to
change drastically. In particular, it is not justified to blindly
identify the extracted mass of the top quark as the pole mass. The
size of this additional error will depend most of all on whether the
analysis is done at LO or NLO. It also depends on what values of
$\mu_{\text{PS}}$ are considered acceptable - increasing the range in
$\mu_{\text{PS}}$ increases the spread in the extracted
mass. We find that the ambiguity in the (indirect) determination of
$\overline{m}$ is smaller than that present in the determination of
the pole mass. This could be due to the better convergence of the
associated perturbative series relating the different masses, 
which in turn is related to the renormalon ambiguities inherent to the 
pole mass. 

\section{Conclusions}
\label{sec:conclusions}

We have presented an approach that allows the computation of cross sections
involving heavy unstable particles, formalising the physical picture
of production and subsequent decay. Our approach makes use of
techniques common to effective-theory calculations, organizing
the amplitude into contributions from matching coefficients and from
dynamic soft degrees of freedom in the effective theory. This split
allows one to disentangle and separately study effects from widely
different scales. Compared to computations using the complex-mass
scheme, our approach leads to calculations that are considerably simpler 
technically. However, our results do have the disadvantage of
being valid only in the resonant region. To relax this constraint
within the effective theory approach, we would have to match our
results to a strict fixed-order calculation (without any self-energy
resummation) outside the resonance region. Similarly, if we wanted to
extend the results to the threshold region, we would have to match our
effective theory to another NRQCD-like effective theory that describes
a top pair near resonance.

The approach applied in this paper to top pair production from initial
state quarks is a generalization and development of an approach
applied earlier to single-top production. In particular, the real
corrections are expanded at the integrand level so that no further
expansions after phase space integration are required. 

Generically, off-shell effects are found to be small and only have an
impact near kinematic edges where distributions are not inclusive
in the invariant mass and, therefore, the corresponding cancellations
are not complete.  As a toy application we have investigated the
impact of mass-scheme ambiguities for the extraction of the top mass
from the invariant mass distribution. In a simple analysis we found
scheme ambiguities of $500-900$~MeV and $400-500$~MeV for the
determination of the pole mass using LO and NLO calculations
respectively. Future applications of the EFT method include the
possibility of studying resummation of logarithms of ratios of widely
separated scales for fully differential cross sections. The fact that
effective theories disentangle effects from different scales is a
promising starting point for such an investigation.

\section{Acknowledgements}
AP gratefully acknowledges the support and hospitality of the CERN and
PSI Theory groups, where part of this work was completed. This
research has been supported, in part, by the UK Science and Technology
Facilities Council and the ERC grant 291377 ``LHCtheory: Theoretical
predictions and analyses of LHC physics: advancing the precision
frontier."  The work of PF is supported by the ``Stichting voor
Fundamenteel Onderzook der Materie (FOM)".

\bibliography{ttbar_nonfact}
  
\end{document}